\documentclass[sigconf,nonacm]{acmart}
\AtBeginDocument{%
  }

\setcopyright{acmlicensed}
\copyrightyear{2018}
\acmYear{2018}
\acmDOI{XXXXXXX.XXXXXXX}
\acmConference[Conference acronym 'XX]{Make sure to enter the correct
  conference title from your rights confirmation email}{June 03--05,
  2018}{Woodstock, NY}
\acmISBN{978-1-4503-XXXX-X/2018/06}

\usepackage[utf8]{inputenc} %
\usepackage[T1]{fontenc}    %
\usepackage{hyperref}       %
\usepackage{url}            %
\usepackage{booktabs}       %
\usepackage{amsfonts}       %
\usepackage{nicefrac}       %
\usepackage{microtype}      %
\usepackage{xcolor}         %
\usepackage{wrapfig}
\usepackage{bm}
\usepackage{xspace}
\usepackage{arydshln} 
\usepackage{colortbl}
\usepackage{subcaption} %
\usepackage{latexsym}
\usepackage{enumitem}
\usepackage{makecell}
\usepackage{threeparttable}
\usepackage{amsmath}
\usepackage{algorithm}
\usepackage{algpseudocode}
\usepackage{float} %
\usepackage{multirow}
\usepackage[most]{tcolorbox}
\newtcolorbox{alprompt}[1]{
        boxrule = 1pt,
        fontupper = \small\tt,
        fonttitle = \bf\color{black},
        arc = 2pt,
        rounded corners,
        colframe = black,
        colbacktitle = white!97!yellow,
        colback = white!97!yellow,
        title = #1,
}

\usepackage{tcolorbox}
\newtcolorbox{findingbox}{
  colback=gray!5,
  colframe=gray!50,
  boxrule=0.5pt,
  arc=2pt,
  left=6pt, right=6pt, top=4pt, bottom=4pt,
}

\usepackage{graphicx}

\begin{document}

\title{ZeroCoder: Can LLMs Improve Code Generation Without Ground-Truth Supervision?}

\author{Lishui Fan}
\affiliation{
  \institution{Zhejiang University}
  \department{The State Key Laboratory of Blockchain and Data Security}
  \country{China}
}
\email{flscode@zju.edu.cn}

\author{Mouxiang Chen}
\affiliation{
  \institution{Zhejiang University}
  \department{The State Key Laboratory of Blockchain and Data Security}
  \country{China}
}
\email{chenmx@zju.edu.cn}

\author{Tingwei Zhu}
\affiliation{
  \institution{Zhejiang University}
  \department{The State Key Laboratory of Blockchain and Data Security}
  \country{China}
}
\email{tingwei.zhu@zju.edu.cn}

\author{Kui Liu}
\affiliation{
  \institution{HuaWei}
  \country{China}
}
\email{brucekuiliu@gmail.com}

\author{Xin Xia}
\affiliation{
  \institution{Zhejiang University}
  \country{China}
}
\email{xin.xia@acm.org}

\author{Shanping Li}
\affiliation{
  \institution{Zhejiang University}
  \department{The State Key Laboratory of Blockchain and Data Security}
  \country{China}
}
\email{shan@zju.edu.cn}

\author{Zhongxin Liu}
\authornote{Corresponding author}
\affiliation{
  \institution{Zhejiang University}
  \department{The State Key Laboratory of Blockchain and Data Security}
  \country{China}
}
\email{liu_zx@zju.edu.cn}

\definecolor{codebg}{HTML}{F8F8F8}
\definecolor{keyword}{HTML}{CF222E}
\definecolor{string}{HTML}{0A3069}
\definecolor{builtin}{HTML}{8250DF}
\definecolor{comment}{HTML}{6E7781}
\lstdefinestyle{pythonstyle}{
  language=Python,
  basicstyle=\ttfamily\scriptsize,
  keywordstyle=\color{keyword}\bfseries,
  stringstyle=\color{string},
  commentstyle=\color{comment}\itshape,
  showstringspaces=false,
  breaklines=true,
  frame=none,
  xleftmargin=0pt,
  aboveskip=2pt,
  belowskip=2pt,
  morekeywords={assert, True, False, None},
}
\newcommand\app{ZeroCoder\xspace}
\newcommand\appdy{\ensuremath{Dy\mathcal{B}^4}\xspace}

\algrenewcommand\algorithmicrequire{\textbf{Input:}}
\algrenewcommand\algorithmicensure{\textbf{Output:}}

\newcommand{\Exec}{\texttt{exec}}
\newcommand{\Rank}{\mathrm{Rank}}
\newcommand{\Sel}{\mathrm{Sel}}
\begin{abstract}

Code generation is an important task in software engineering, and Reinforcement Learning with Verifiable Rewards (RLVR) has emerged as a powerful paradigm for improving it through execution-based feedback. 
However, most RLVR pipelines rely on human-curated unit tests as verifiers, making further progress bottlenecked by the cost and scarcity of supervision. 
Existing work has tried to use self-generated tests to ground rewards, but the effect is constrained by the lack of discriminative tests due to the sub-optimal performance of the model on test generation.
Our idea is to improve code generation without ground-truth supervision by co-evolving both code and test generation, so that their interactions produce progressively more informative supervision.
To this end, we present \app, a fully label-free co-evolutionary framework that jointly trains a Coder, which generates solutions, and a Tester, that generates tests, using only execution feedback from self-generated code–test interactions.  
For each problem, \app executes sampled solutions against sampled tests to form a passing matrix, identifies a consensus subset of likely-correct solutions and consistent tests via a pluggable selection algorithm, and derives role-specific rewards accordingly. To ensure reward quality, \app filters low-information instances via rank-based pre-filtering and trains the Tester with a curriculum balancing validity and mutation-driven discriminativeness.
We further identify selector drift, the progressive miscalibration of fixed selection rules during co-evolution, and introduce \appdy, a Bayesian selector that uses as few as 10 labeled instances to dynamically recalibrate its priors.
Across three model families and six benchmarks, \app consistently improves both code generation and test generation. In the fully label-free setting, \app yields strong gains. For example, on Qwen2.5-Coder-7B-Instruct, it improves code generation by up to 14.5\% over the base model. With \appdy, the gain reaches 21.6\%, while test generation improves by 24.3\%, approaching oracle-supervised performance.
\end{abstract}

\received{20 February 2007}
\received[revised]{12 March 2009}
\received[accepted]{5 June 2009}

\maketitle

\section{Introduction}
Code generation is an important task in software engineering~\cite{zhang2025unseen,yu2024codereval}.  Reinforcement Learning with Verifiable Rewards (RLVR) has emerged as a dominant paradigm for improving it, delivering substantial gains by optimizing large language models (LLMs) against execution-based feedback~\cite{guo2025deepseek,yang2025qwen3}. Yet this success still depends heavily on the availability of ground-truth supervision: in practice, RLVR pipelines typically rely either on human-curated unit tests or on tests automatically generated from reference solutions as verifiers~\cite {ma2025unitcoder,du2023classeval,liu2024your}, 
both requiring substantial manual effort and high-quality guarantees.
As these resources become increasingly expensive to scale~\cite{du2023classeval,wang2023execution}, further progress is bottlenecked by supervision scarcity. This raises a pressing question: \textit{can LLMs improve code generation without relying on ground-truth supervision?}

A growing line of work has explored self-rewarding mechanisms to improve LLM reasoning without external labels, where models generate their own feedback to serve as a training signal. One line constructs rewards from model-side self-assessment signals, such as confidence or entropy~\cite{jang2025self,zhang2025right,prabhudesai2025maximizing}; another derives supervision from consensus-based selection over multiple self-generated candidates, such as majority voting~\cite{zuo2025ttrl,huang2025r}. Although different in form, both depend on internal model agreement as a proxy for correctness, which may reinforce spurious yet self-consistent trajectories and even induce performance collapse~\cite{shafayat2025can}.
For code generation, a more objective proxy is program execution. Rather than relying on internal model agreement, one can assess solution quality by whether the code actually runs correctly. Along this direction, recent work uses self-generated tests and execution feedback to evaluate candidate programs~\cite{chen2025self,wang2025co}. However, this route remains fundamentally bottlenecked by the quality of self-generated tests: due to the inherent scarcity of high-quality tests compared to code in training corpora~\citep{lin2025learning,prasad2025learning}, self-generated tests often lack discriminative power, allowing spurious solutions to pass and yielding noisy, low-information reward signals.
Consequently, in the absence of ground-truth supervision, improving code generation requires strengthening the model's test-generation capability.

Since code generation and test generation are dual tasks relying on a shared understanding of the code problem~\cite{xiong2024program,chencodet}, our idea is to \textit{co-evolve} the two tasks during reinforcement learning. A single model alternates between two roles, acting as a coder to produce candidate solutions and as a tester to produce candidate tests. 
Rather than treating one role as fixed while optimizing the other, both roles are jointly trained and mutually improve through role-specific reward signals derived from the resulting code-test interactions: stronger tests place greater pressure on the coder to produce correct solutions, and a stronger coder, in turn, raises the bar for the tester to generate more discriminative tests. 
Crucially, in our label-free setting, these code-test interactions are the only source of supervision. The success of co-evolution, therefore, depends on whether they can yield informative reward signals for both roles.
Realizing this idea requires addressing two challenges. 
From the training data perspective, problems that are too easy or too hard lead to degenerate code-test interactions with nearly identical rewards (e.g, either all pass or fail), driving group-normalized advantages toward zero and resulting in vanishing policy gradients that destabilize training~\cite{yu2025dapo,yang2025qwen3}.
From the training process perspective, since the model's ability as a tester is initially weaker than as a coder, naively optimizing both roles together may fail to bootstrap the tester into a reliable and discriminative verifier.

In this paper, we propose \app, a fully label-free framework that improves code generation by learning from self-generated code-test interactions, without relying on any ground-truth supervision. 
For each problem, \app samples candidate solutions from the coder and tests from the tester, and executes them to form a passing matrix that summarizes pairwise execution outcomes.
Based on this matrix, a selection algorithm, which we refer to as a \textit{selector}, identifies a consensus subset of likely-correct solutions and consistent tests, i.e., a subset in which solutions exhibit mutually consistent execution behavior with respect to the selected tests, and \app derives role-specific rewards for both the coder and the tester. 
To address the two challenges above, \app incorporates two mechanisms. 
First, \app performs an offline rank-based pre-filtering step before training, retaining only problems whose passing matrices exhibit sufficient interaction diversity measured by matrix rank. Second, \app introduces a mutation-based reward for tester training, which is integrated into a curriculum designed to balance validity and discriminativeness. This curriculum rewards tests for agreeing with selector-induced consensus solutions, and gradually shifts emphasis toward mutation-based discriminativeness by encouraging tests to expose subtly faulty implementations.

Built on \app, we instantiate three representative selectors, namely MaxPass~\citep{li2022competition}, which selects solutions with the highest pass counts, CodeT~\citep{chencodet}, which leverages agreement between generated solutions and tests, and $\mathcal{B}^4$~\citep{chen2024b4}, which ranks candidate solutions using Bayesian posterior estimates derived from execution outcomes. We find that \app already improves both code generation and test generation in a fully label-free setting.
However, through further analysis, we identify the \textit{selector drift} phenomenon: existing selectors encode fixed inductive biases (e.g., MaxPass assumes sufficiently reliable tests~\citep{chen2024b4}), which may become miscalibrated as the reliability of the coder and tester drifts during co-evolutionary training, eventually yielding increasingly noisy reward signals.
To mitigate this issue, the selector should be recalibrated as training progresses. We therefore propose Dynamic $\mathcal{B}^4$ (\appdy), a dynamically calibrated variant of $\mathcal{B}^4$. Whereas the original $\mathcal{B}^4$ relies on fixed prior assumptions when forming posterior estimates, \appdy updates its prior hyperparameters using a very small labeled calibration set (e.g., as few as 10 labeled instances) to maintain selector quality throughout training.

Extensive evaluations demonstrate the effectiveness of \app and \appdy across four code-generation and two test-generation benchmarks on three models (i.e., Qwen2.5-1.5B-Instruct~\cite{hui2024qwen2}, Qwen3-4B~\cite{yang2025qwen3}, and Qwen2.5-Coder-7B-Instruct~\cite{yang2025qwen2}). Taking Qwen2.5-Coder-7B-Instruct as an example, in the fully label-free setting, \app with the selector $\mathcal{B}^4$~\cite{chen2024b4} achieves 14.5\% and 15.6\% relative improvements over the base model in code generation and test generation, respectively. Equipping \app with \appdy yields further improvements of 21.6\% and 24.3\%, which is competitive with oracle training that relies on ground-truth supervision.

In summary, our main contributions are as follows:

\begin{itemize}[left=0em, topsep=0pt]

\item \textbf{\app}. A label-free co-evolutionary RLVR framework that jointly improves code/test generation from self-generated code-test interactions. By improving the informativeness of self-generated rewards through rank-based pre-filtering and discriminative tester optimization, \app delivers consistent gains without ground-truth supervision.

\item \textbf{Selector Drift Analysis \& \appdy}. We identify selector drift, the progressive miscalibration of fixed selection rules under co-evolution, as a key source of reward noise. We propose \appdy, a dynamically calibrated Bayesian selector that uses a small labeled set to maintain selection quality throughout training.

\item \textbf{Empirical Validation \& Analysis}. On average across three representative models and six benchmarks, \app with \appdy improves code generation and test generation by 18.8\% and 62.7\% over the base models, respectively, matching oracle-supervised performance. Further analysis confirms that both the offline-filtering and mutation-driven discriminativeness reward components are critical.

\end{itemize}

\section{Preliminaries}
\label{sec:preliminaries}
\paragraph{Problem Formulation.}
We study label-free training for code generation and test generation. We employ a single language model $\pi_\theta$, which alternates between two roles via prompting during training: a coder $\pi_\theta^{\text{coder}}$ and a tester $\pi_\theta^{\text{tester}}$.  

Given a problem $x$, the coder samples $N$ candidate solutions $\mathcal{S}=\{s_i\}_{i=1}^{N}$ with $s_i\sim\pi_\theta^{\text{coder}}(\cdot\mid x)$ and the tester samples $M$ candidate tests  $\mathcal{T}=\{t_j\}_{j=1}^{M}$ with $t_j\sim\pi_\theta^{\text{tester}}(\cdot\mid x)$.
The interaction between $\mathcal{S}$ and $\mathcal{T}$ is captured by executing every pair $(s_i, t_j)$.
We define the execution result as $e_{ij} = \mathbb{I}(\text{exec}(s_i, t_j) = \text{pass})$, where $\mathbb{I}(\cdot)$ is the indicator function. 
The aggregation of these outcomes forms a passing matrix $E=\{e_{ij}\}_{N\times M} \in \{0, 1\}^{N \times M}$, where the entry $e_{ij}$ indicates whether solution $s_i$ passes test $t_j$. 

A selector $\mathrm{Sel}$ maps $E$ to a selected consensus subset of solutions and tests:
$$
\mathrm{Sel}(E)=(\mathcal{C}_S,\mathcal{C}_T),\quad
\mathcal{C}_S\subseteq \mathcal{S},\ \mathcal{C}_T\subseteq \mathcal{T}
$$
In this paper, we consider representative selectors, including MaxPass, CodeT, and $\mathcal{B}^4$, which differ in how they select $\mathcal{C}_S$ and $\mathcal{C}_T$ from the execution matrix $E$.
In particular, we primarily use $\mathcal{C}_S$ as a proxy set of higher-quality solutions for reward construction.

\paragraph{Reinforcement Learning with Verifiable Rewards (RLVR)}
Standard RLVR for code generation assumes a supervised dataset $\mathcal{D}_{\text{sup}} = \{(x, T^*)\}$, where $T^*$ represents ground-truth verification resources (e.g., test cases).
It optimizes
$$
\mathcal{J}_{\text{RLVR}}(\theta)
=\mathbb{E}_{x\sim\mathcal{D}_{\text{sup}},\, s\sim\pi_\theta(\cdot\mid x)}
\big[r(s,T^*)\big]
$$
where $r(s, T^*)$ is the objective reward (e.g., $r=1$ if $s$ passes all tests in $T^*$, else 0).
In the label-free setting, we derive learning signals from the model's own code-test interactions, eliminating the reliance on large-scale, reliable verifiers.

\section{\app}

\begin{figure*}[t]
    \centering
    \includegraphics[width=0.95\linewidth]{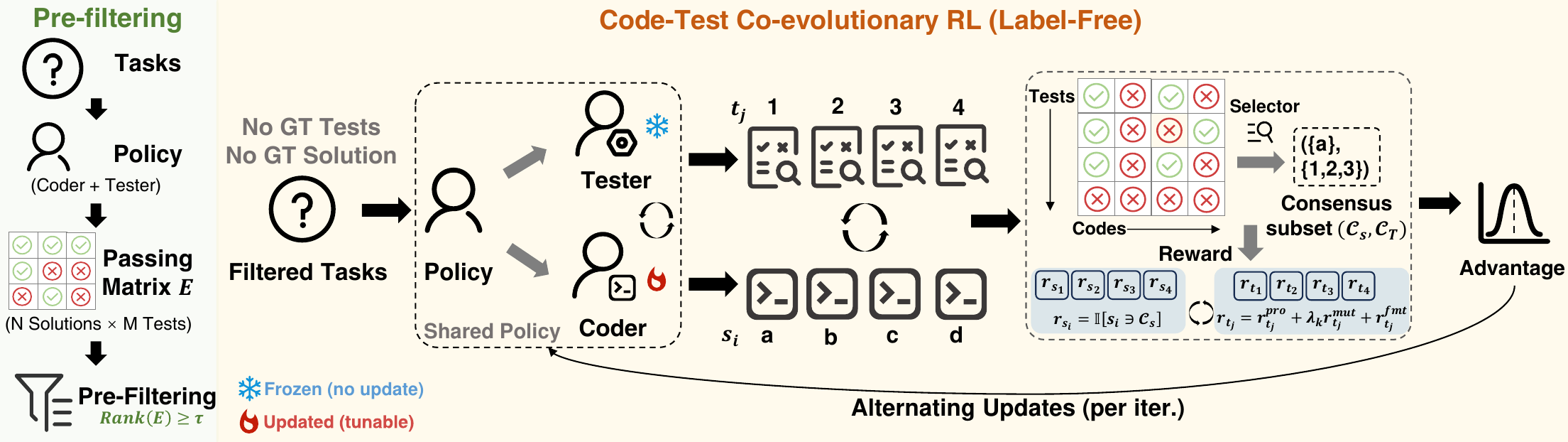}
\caption{Overview of \app framework.}
    \label{fig:main-figure}
\end{figure*}

\app co-trains a coder and a tester without ground-truth solutions or tests, using execution feedback from self-generated interactions. Making such co-evolution work requires addressing two key challenges: (i) problems that are too easy or too hard for the current policy induce degenerate passing matrices and uninformative rewards, and (ii) the tester is initially weak, and tends to produce tests that lack discriminativeness. To address these challenges, \app comprises two components: offline rank-based pre-filtering (\S\ref{sec:offline}), which removes low-information problems before RL, and code-test co-evolutionary RL (\S\ref{sec:coevo}), which jointly optimizes both roles under execution feedback while equipping the tester with a curriculum-based objective that progressively improves its validity and discriminativeness. We further introduce Dynamic $\mathcal{B}^4$ (\S\ref{sec:dyb4}) as an optional selector that can be integrated into \app to mitigate selector drift during co-evolution. The overview of \app is shown in Figure~\ref{fig:main-figure}.

\begin{figure}[t]
    \centering
    \includegraphics[width=0.9\linewidth]{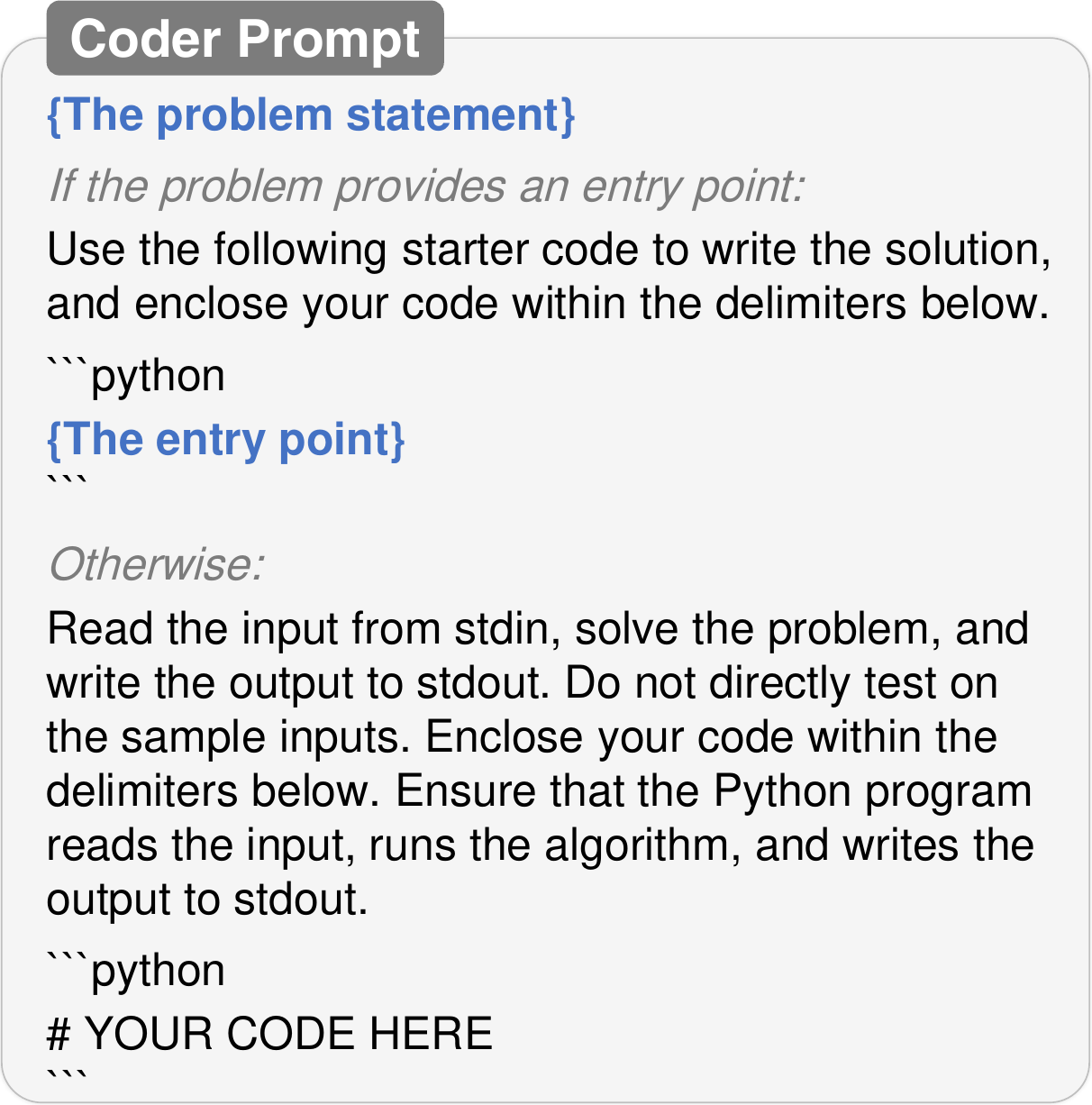}
    \caption{Coder prompt template used in our experiments.}
    \label{fig:coder-prompt}
\end{figure}

\subsection{Offline Rank-based Pre-filtering}
\label{sec:offline}

Effective RL benefits from training instances that induce informative reward variation across candidates~\citep{zhao2025absolute,yu2025dapo}. 
Existing RLVR methods~\citep{wen2025light,guo2025deepseek,yang2025qwen3} often rely on accuracy-based filtering (e.g., retaining instances with non-trivial pass rates) to preserve informative reward variation. In such supervised settings, an instance supports informative selection only when it contains a mix of correct and incorrect candidates, providing the necessary contrast for the algorithm to distinguish superior solutions from inferior ones.

In our label-free setting, however, ground-truth accuracy is unavailable. 
The only observable supervision comes from the passing matrix $E$, which records the outcomes of interactions between self-generated solutions and self-generated tests. 
Therefore, whether an instance is informative depends on the diversity of interaction patterns encoded in $E$: informative instances should contain solutions with different pass/fail signatures across tests, and tests that distinguish solutions in different ways.

When this diversity collapses, reward construction becomes degenerate. 
If $E$ is nearly all ones, most tests are too weak to distinguish solutions; if $E$ is nearly all zeros, almost all solutions fail and little actionable feedback is available. 
More generally, whenever many rows or columns of $E$ are redundant, the resulting rewards become nearly constant across candidates, leading to near-zero group-normalized advantages~\citep{yu2025dapo} and thus weak policy gradients.

This motivates the need for an unsupervised proxy computed solely from $E$. 
We use the rank of $E$ as an unsupervised proxy for interaction diversity~\citep{koren2009matrix,roy2007effective}. 
Intuitively, rank reflects how many non-redundant execution signatures are present in the matrix: low rank implies many solutions behave similarly or many tests are redundant, whereas a higher rank indicates a richer interaction structure for more informative selection and reward construction.

Based on this insight, we curate a high-information dataset $\mathcal{D}$ from the raw dataset $\mathcal{D'}$.
Before training, we use the initial policy $\pi_{\theta_0}$ to perform a one-time rollout for each problem $x \in \mathcal{D'}$ to construct the matrix $E_x$.
We then compute the rank $\mathrm{Rank}(E_x)$ and retain only those problems whose interaction diversity exceeds a threshold $\tau$, yielding
$$
\mathcal{D} = \{ x \in \mathcal{D'} \mid \mathrm{Rank}(E_x) \ge \tau \}
$$
This filtering removes low-information instances and ensures that the subsequent RL phase focuses on samples with high potential for co-evolutionary gain.

\subsection{Code-Test Co-evolutionary RL}

\label{sec:coevo}
We optimize $\pi_\theta$ by converting the passing matrix $E$ into role-specific rewards through a selector-agnostic reward-construction module. 
Given $E$, the module (i) applies a selector $\mathrm{Sel}$ to obtain a consensus subset of solutions and tests, denoted by $\mathcal{C}_S$ and $\mathcal{C}_T$, and (ii) assigns rewards to the coder and the tester based on $(E,\mathcal{C}_S)$.
Notably, it is selector-agnostic: different selectors can be plugged in, and a single selector is kept fixed within each training run.

We instantiate three representative selectors: MaxPass~\citep{li2022competition}, CodeT~\citep{chencodet}, and $\mathcal{B}^4$~\citep{chen2024b4}.
MaxPass ranks solutions by the number of passed tests, assuming that the sampled tests contain sufficiently reliable cases to separate correct from spurious solutions.
CodeT partitions solutions into functionality-equivalent groups based on their execution signatures on the sampled tests. Specifically, solutions that exhibit identical pass/fail patterns across tests are grouped together. CodeT then scores each group according to the strength of its consensus and selects the highest-scoring group as the consensus group, assuming correct solutions appear with non-trivial probability so that a high-quality group can emerge.
$\mathcal{B}^4$~\cite{chen2024b4} denotes a specific instantiation of the Bayesian selector with two fixed prior hyperparameters, $\beta_0$ and $\alpha_{xy}$. It scores candidate correctness configurations, each of which induces a candidate consensus set $(\tilde{\mathcal{C}}_S,\tilde{\mathcal{C}}_T)$. Following~\cite{chen2024b4}, its practical score is approximated as
$$
\mathrm{score}_{\mathcal{B}^4}
  (\tilde{\mathcal{C}}_S, \tilde{\mathcal{C}}_T;\, \beta_0, \alpha_{xy})
=
\prod_{k \in \{1,0,x,y\}}
  \mathrm{B}(a_k, b_k)
$$
where each $(a_k, b_k)$ combines observed summary statistics from $E$ with fixed prior hyperparameters $\boldsymbol{\eta}=(\beta_0, \alpha_{xy})$: $\beta_0$ encodes the belief that incorrect solutions rarely pass incorrect tests, and $\alpha_{xy}$ favors larger consensus sets. $\mathcal{B}^4$ returns the $\tilde{\mathcal{C}}_S$ with the highest score. See~\cite{chen2024b4} for the full formulation.

These assumptions are reasonable under fixed operating conditions, but may become miscalibrated as the coder/tester's reliability drifts during training, motivating \appdy (\S\ref{sec:dyb4}). For convenience, we refer to these selectors as \textit{static selectors}, since their decision rules remain fixed throughout training. 

\noindent\textbf{Coder Reward.}
In the absence of ground-truth solutions, we treat the selected solutions $\mathcal{C}_S$ as a proxy for higher-quality solutions under the current interaction signal.
We assign a binary reward to solutions selected into $\mathcal{C}_S$:
$$
r_{s_i} = \mathbb{I}\big[s_i \in \mathcal{C}_S\big]
$$
where $\mathbb{I}(\cdot)$ is the indicator function.

\noindent\textbf{Tester Reward.}
Optimizing the tester requires balancing validity with respect to likely correct solutions and discriminative power. Since ground-truth labels are absent, we leverage $\mathcal{C}_S$ as a proxy oracle while explicitly discouraging trivial tests. We design the overall tester reward to consist of three different terms:

\textit{Proxy-Agreement Term.}
We reward a test if it is satisfied by a proxy-good solution.
To reduce training-time overhead, instead of evaluating a candidate test $t_j$ on all solutions in $\mathcal{C}_S$, we randomly sample one $s^\star\sim\mathcal{C}_S$ and define
$$
r^{\text{pro}}_{t_j} = \mathbb{I}\big(\texttt{exec}(s^\star, t_j)=\texttt{pass}\big)
$$

\textit{Discriminative Term.}
Crucially, the tester can satisfy $r^{\text{pro}}$ by generating trivial tests (e.g.,  \texttt{assert True}), yielding little useful signal for learning.
To mitigate this, we introduce a mutation-based score based on program mutation testing. 
We apply a program mutation tool\footnote{In our implementation, we use the \texttt{mutmut} library.} to generate a set of mutated variants of $s^\star$, denoted by $\mathcal{M}(s^\star)$, and measure how many mutants are killed by $t_j$:

$$
r^{\text{mut}}_{t_j}
=\frac{\sum_{m\in\mathcal{M}(s^\star)}\mathbb{I}\big(\texttt{exec}(m,t_j)=\texttt{fail}\big)}
{|\mathcal{M}(s^\star)|}
$$
Even if a test aligns with the consensus, it receives a low mutation reward if it cannot distinguish $s^\star$ from its mutants, discouraging low-discriminability tests.

\textit{Format Term.}
To ensure tests are parsable, we penalize unparsable outputs:
$$
r_{t_j}^{\text{fmt}} =
\begin{cases}
0, & \text{if $t_j$ is executable},\\
-1, & \text{otherwise}.
\end{cases}
$$

\textit{Overall tester Reward.}
We combine the terms as:
$$
r_{t_j} = r^{\text{pro}}_{t_j} + \lambda_k \, r_{t_j}^{{\text{mut}}} + r_{t_j}^{\text{fmt}}
$$
We use a scheduler $\lambda_k$ that gradually upweights discriminativeness.
Let $\texttt{progress}=\frac{k}{K}\in[0,1]$ be the normalized training step, where $k$ is the current step and $K$ is the total number of RL steps. We set
$$
\lambda_k = 1.25 - 0.25  \cos(\pi \cdot \texttt{progress})
$$
so $\lambda_k$ increases smoothly from $1.0$ to $1.5$.

This curriculum on $\lambda_k$ reflects a developmental progression from validity to discriminative power.
Early in training, the tester is relatively weak and should first learn to produce valid tests for likely correct solutions. We therefore avoid over-emphasizing the mutation term $r^{\text{mut}}$ at the beginning, and gradually upweight it as the tester becomes more reliable, so that training shifts from encouraging validity to encouraging finer-grained discrimination.
We adopt a cosine schedule, a commonly used schedule~\citep{loshchilov2017sgdr,nichol2021improved}, to ensure a smooth transition from validity-focused training to discriminativeness-focused training.

Beyond reward design, we also adopt an alternating update scheme to reduce non-stationarity. Specifically, we alternate between coder and tester updates across training steps, rather than updating both roles sequentially within the same step. The latter would make the second update depend on off-policy data after the first role has changed, which may increase non-stationarity and degrade training quality~\cite{shao2024deepseekmath,zawalski2021off}.

\begin{figure}[t]
    \centering
    \includegraphics[width=0.9\linewidth]{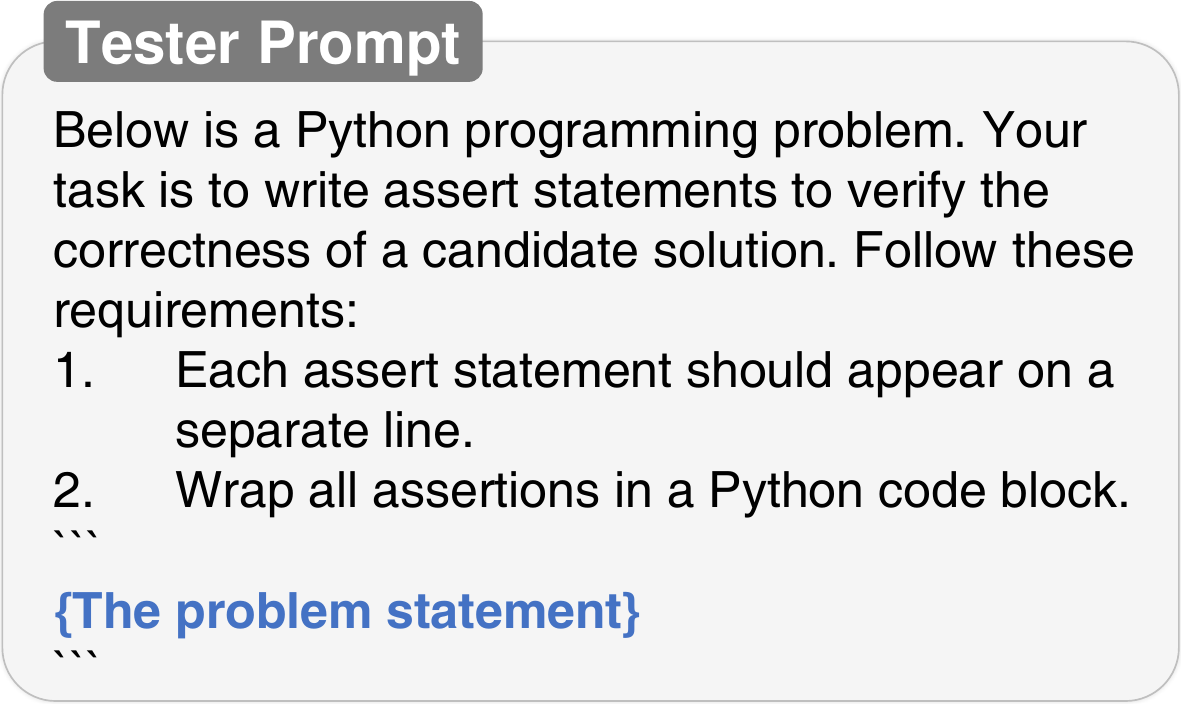}
    \caption{Tester prompt template used in our experiments.}
    \label{fig:tester-prompt}
    \vspace{-1.4em}
\end{figure}

\subsection{Dynamic $\mathcal{B}^4$}
\label{sec:dyb4}
While \app itself is label-free, static selectors may become miscalibrated as the reliability of the coder and tester evolves during training (see \S\ref{sec:base_result}). Such miscalibration can lead to noisier consensus selection and, consequently, noisier reward signals. To mitigate this effect, we propose Dynamic $\mathcal{B}^4$ (\appdy), a label-assisted selector built on top of the original selector $\mathcal{B}^4$, which recalibrates $\mathcal{B}^4$'s prior hyperparameters to align with the current policy's capability.

We construct a lightweight labeled set $\mathcal{H}=\{(x,s^\dagger, T^\dagger)\}$ for selector calibration, where each instance contains a problem $x$, a reference solution $s^\dagger$, reference tests $T^\dagger$. 
We denote the hyperparameters by $\bm{\eta}=(\beta_0,\alpha_{xy})$.  Following $\mathcal{B}^4$~\citep{chen2024b4}, we define a candidate search space $\Omega$ as a log-scale grid with $\beta_0\in[10^{0},10^{9}]$ and $\alpha_{xy}\in[10^{0},10^{7}]$, yielding a search space of $10 \times 8 = 80$ candidate configurations.
Before each training step $k$, we use the current policy to sample candidate solutions and tests for each problem in $\mathcal{H}$, construct passing matrices, and run $\mathcal{B}^4$ under each $\bm{\eta}\in\Omega$.
Let $\hat{s}_{\bm{\eta}}(x)$ be the solution selected by $\mathcal{B}^4$ with hyperparameters $\bm{\eta}$ for calibration problem $x$.
We then choose:
\[
\bm{\eta}^{(k)}
=\arg\max_{\bm{\eta}\in\Omega}
\frac{1}{|\mathcal{H}|}\sum_{(x,s^\dagger,T^\dagger)\in\mathcal{H}}
\mathbb{I}\!\left(\texttt{exec}(\hat{s}_{\bm{\eta}}(x),T^\dagger)=\texttt{pass}\right)
\]
and use $\bm{\eta}^{(k)}$ for $\mathcal{B}^4$-based reward construction at step $k$.
The cost is manageable, as $|\mathcal{H}|=10$ already achieves comparable calibration quality to substantially larger calibration sets (see \S\ref{sec:abl_cal}), and the calibration step accounts for only 2\% of the per-step wall-clock time (see \S\ref{sec:rq3_analysis}).

\section{Experiment Setup}
\paragraph{Models.}
We conduct training across different model scales, including Qwen2.5-1.5B-Instruct~\cite{yang2025qwen2}, Qwen3-4B~\citep{yang2025qwen3} and Qwen2.5-Coder-7B-Instruct~\citep{hui2024qwen2}.
Due to resource constraints, we run Qwen3-4B in its non-thinking mode throughout training and evaluation.

\paragraph{Benchmark and Metrics.}
We select four widely used code-generation benchmarks, including MBPP~\citep{austin2021program}, LiveCodeBench~\citep{jain2025livecodebench}, APPS~\citep{hendrycks2measuring} and CodeForces~\citep{penedo2025codeforces}.
For LiveCodeBench, we follow prior work~\citep{yang2025qwen3} and use problems from Oct 2024 to Feb 2025 to reduce potential data leakage. For computational efficiency on large-scale benchmarks, we evaluate on subsets: following prior practice~\citep{olausson2023self,lecodechain}, we randomly sample 300 problems from the APPS test set, preserving its difficulty distribution (60/180/60 for introductory/interview/competition). For CodeForces, we randomly sample 300 problems and ensure this subset has no overlap with the APPS problems used in our experiments. We report Pass@1 under greedy decoding to ensure reproducibility.

To quantify test generation capability, we evaluate on LiveCodeBench and MBPP. For each problem, we generate tests via greedy decoding and assess them using a fixed reference solution.
(i) \emph{Test Accuracy} is measured by the pass rate of generated tests $T$ on the reference solution $s^{ref}$:

$$
\small{
\texttt{ACC}(T) \;=\; \frac{1}{|T|}\sum_{t\in T} \mathbb{I}\!\left(\texttt{exec}(s^{ref}, t)=\texttt{pass}\right)
}
$$
(ii) \emph{Mutation score} measures the average mutant-killing ability of generated tests. We use \texttt{mutmut} to generate a mutant set $\mathcal{M}(s^\star)$, compute for each generated test the fraction of mutants it kills, and then average this quantity over all generated tests.
$$
\small{
\texttt{Mut}(T)
=
\frac{1}{|T|\,|\mathcal{M}(s^{ref})|}
\sum_{t\in T}\sum_{m\in\mathcal{M}(s^{ref})}
\mathbb{I}\!\left(\texttt{exec}(m,t)=\texttt{fail}\right)
}
$$ 

Since LiveCodeBench lacks ground-truth solutions, following prior practice~\citep{wang2025co}, we construct the pseudo ground-truth solution set by sampling 8 solutions using Qwen3-235B-A22B~\citep{yang2025qwen3} with temperature $=0.8$ and selecting one solution that passes all official tests. To ensure compatibility with mutation tools, we restrict to function-style problems where the solution contains a clear entry point. This choice is also consistent with prior practice~~\citep{wang2025co,laban2025llms,xu2025kodcode}. Applying this filtering yields 323 LiveCodeBench instances with pseudo ground-truth solutions for test evaluation. For MBPP, we use the canonical reference solution provided by the benchmark.

\paragraph{Baselines.}
We compare against the following baselines:
(1) Base Model (No RL). The original model without training. This is used to demonstrate the overall performance gains achieved by our framework.
(2) Offline Test-Driven RL. For each training problem, we first generate a fixed set of synthetic tests $t_{\text{syn}}$ using the base model with greedy decoding to ensure reproducibility.
We then train the model only as a coder. The generated codes are executed on the fixed synthetic tests with a binary reward
$
r(s)=\mathbb{I}\!\left(\forall t\in T_{\text{syn}},\ \texttt{exec}(s,t)=\texttt{pass}\right)
$, where $T_{\text{syn}}$ is the fixed synthetic tests. It is designed to reveal the limitations of relying on static pseudo-labels.
(3) Online Test-Driven RL. During training, we generate tests online and train the model only as a coder. We use a binary reward that is 1 iff the code passes all sampled tests; otherwise, it is 0:
$
r(s)=\mathbb{I}\!\left(\forall t\in T_{\text{sam}},\ \texttt{exec}(s,t)=\texttt{pass}\right)
$, where $T_{\text{sam}}$ is the sampled tests.
Comparing \app against this baseline explicitly isolates and validates the necessity of co-evolving the tester.
For a fair comparison, both RL baselines use the same sampling configuration during training as \app.

\paragraph{Training Details}
We optimize the policy using the GRPO objective~\citep{guo2025deepseek} and implement training in VeRL~\citep{sheng2025hybridflow} on the training split of the APPS dataset~\citep{hendrycksapps2021}, which contains 5,000 coding problems. For the static selector $\mathcal{B}^4$, we adopt two settings, $\mathcal{B}^4(4,3)$ and $\mathcal{B}^4(5,3)$, as used in prior work~\citep{chen2024b4}. 
The threshold $\tau$ for filtering low-information instances is set to 2. 
After pre-filtering, the resulting training sets contain 1,377, 1,396 and 990 instances for  Qwen2.5-1.5B-Instruct, Qwen3-4B and Qwen2.5-Coder-7B-Instruct, respectively.  The prompts for coder and tester are shown in Figure~\ref{fig:coder-prompt} and~\ref{fig:tester-prompt}, respectively.
For \appdy, we reserve a labeled set of size $|\mathcal{H}|=10$, where $\mathcal{H}$ is randomly sampled from the filtered APPS training set; the remaining instances are used for training.
We set the actor clipping ratio to 0.2, the KL coefficient to 0.001. All models are trained for 150 steps with a batch size of 32 and a constant learning rate of $1 \times 10^{-6}$.
To balance rollout diversity and computational efficiency, we sample 8 rollouts for both code and test generation at each step, using a temperature of 0.7 and top-$p$ of 0.8. The maximum response length is 2048 tokens. All experiments are conducted on a cluster of 4 NVIDIA A100 GPUs.

\section{Experiments}

In this section, we report and analyze the experimental results to answer the following research questions (RQs):

\begin{itemize}[leftmargin=*]
\item RQ1: How effective is \app in a fully label-free setting for improving code generation and test generation?
\item RQ2: Under a matched small-label setting, how much additional benefit does \app gain from dynamic calibration with \appdy?
\item RQ3: How do \app’s components contribute to its performance, and how does it behave in terms of sensitivity, efficiency, and comparison to oracle-supervised training?
\end{itemize}

We distinguish two evaluation settings throughout the experiments. In the \textit{label-free \app} setting (\S\ref{sec:base_result}), \app is instantiated with static selectors and trained without any labeled data. In the \textit{\app+\appdy} setting (\S\ref{sec:appdy}), we switch to a matched setting with a small amount of labeled data, where the same labeled set $\mathcal{H}$ is introduced for all selectors. Static selectors use it for supervised reward, while \appdy additionally uses it for calibration. 
We then conduct further analysis through component ablations, sensitivity analysis, efficiency profiling, and comparison against oracle-supervised training, to better understand its behavior and contextualize the gains (\S\ref{sec:rq3_analysis}).

\subsection{RQ1: Label-Free \app with Static Selectors}
\label{sec:base_result}
We answer RQ1 by evaluating \app in a fully label-free setting with representative static selectors.
\begin{table*}[t]
\centering
\footnotesize
\caption{Performance of ZeroCoder with static selectors and baseline methods on code- and test-generation benchmarks across three base models in a fully label-free setting. ``LCB'' refers to LiveCodeBench, and ``CF.'' refers to CodeForces. CAvg and TAvg denote the average over code-generation and test-generation benchmarks, respectively. Avg $=$ (CAvg $+$ TAvg) $/$ 2.}
\label{tab:label-free}
\resizebox{0.9\textwidth}{!}{%
\begin{tabular}{@{}ll ccccccc cccc c c@{}}
\toprule
\multirow{3}{*}{\textbf{Model}} & \multirow{3}{*}{\textbf{Selector}} 
& \multicolumn{7}{c}{\textbf{Code Generation}} 
& \multicolumn{4}{c}{\textbf{Test Generation}} 
& \multirow{3}{*}{\textbf{TAvg}} 
& \multirow{3}{*}{\textbf{Avg}} \\
\cmidrule(lr){3-9} \cmidrule(lr){10-13}
& & \multirow{2}{*}{LCB} & \multirow{2}{*}{MBPP} & \multicolumn{3}{c}{Apps} & \multirow{2}{*}{CF.} & \multirow{2}{*}{CAvg} 
& \multicolumn{2}{c}{LCB} & \multicolumn{2}{c}{MBPP} & & \\
\cmidrule(lr){5-7} \cmidrule(lr){10-11} \cmidrule(lr){12-13}
& & & & Intro. & Inter. & Comp. & & & ACC. & Mut. & ACC. & Mut. & & \\
\midrule

\multicolumn{15}{@{}l}{\cellcolor{gray!8}\textit{Qwen2.5-1.5B-Instruct}} \\[2pt]

Base Model & -- 
  & 4.8 & 60.3 & 31.7 & 6.1 & 3.3 & 1.7 & 18.0 
  & 17.5 & 3.0 & 42.7 & 7.2 & 17.6 & 17.8 \\

\quad+Offline & -- 
  & 4.8 & 57.4 & 33.3 & 5.6 & 0.0 & 1.7 & 17.1 
  & 46.1 & 4.3 & 43.9 & 4.5 & 24.7 & 20.9 \\

\quad+Online & -- 
  & 4.8 & 59.3 & 28.3 & 5.0 & 0.0 & 2.0 & 16.6 
  & 52.9 & 9.3 & 45.8 & 6.9 & 28.7 & 22.6 \\

\quad+ZeroCoder 
  & MaxPass 
  & 4.8 & 63.0 & 33.3 & 6.7 & 3.3 & 1.7 & 18.8 
  & 64.7 & 16.7 & 69.0 & 29.9 & 45.1 & 31.9 \\

  & CodeT 
  & \textbf{5.4} & \textbf{64.8} & \textbf{36.7} & \textbf{7.2} & \textbf{5.0} & \textbf{2.3} & \textbf{20.2} 
  & 67.2 & 16.4 & \textbf{71.7} & 24.6 & 45.0 & \textbf{32.6} \\

  & $\mathcal{B}^4$(4,3) 
  & 4.2 & 64.3 & 35.0 & 6.7 & \textbf{5.0} & 1.7 & 19.5 
  & \textbf{70.6} & 15.5 & 67.2 & 27.8 & 45.3 & 32.4 \\

  & $\mathcal{B}^4$(5,3) 
  & \textbf{5.4} & 64.0 & 33.3 & 6.1 & 3.3 & 1.7 & 19.0 
  & 66.9 & \textbf{17.0} & 66.9 & \textbf{32.3} & \textbf{45.8} & 32.4 \\

\midrule

\multicolumn{15}{@{}l}{\cellcolor{gray!8}\textit{Qwen3-4B}} \\[2pt]

Base Model & -- 
  & 21.7 & 79.9 & 50.0 & 21.1 & 1.7 & 8.3 & 30.4 
  & 99.4 & 9.3 & 72.4 & 18.0 & 49.8 & 40.1 \\

\quad+Offline & -- 
  & 22.3 & 75.7 & 50.0 & 18.9 & 0.0 & 10.7 & 29.6 
  & 57.7 & 3.7 & 66.8 & 13.7 & 35.5 & 32.5 \\

\quad+Online & -- 
  & \textbf{24.1} & 79.1 & 46.7 & 19.4 & 0.0 & 12.3 & 30.3 
  & 81.9 & 7.9 & 59.4 & 12.2 & 40.4 & 35.3 \\

\quad+ZeroCoder 
  & MaxPass 
  & 22.9 & 78.8 & 55.0 & 28.9 & 3.3 & 13.0 & 33.6 
  & \textbf{99.7} & 9.6 & 73.3 & 17.2 & 49.9 & 41.8 \\

  & CodeT 
  & \textbf{24.1} & \textbf{80.2} & 56.7 & 27.2 & \textbf{5.0} & \textbf{14.7} & \textbf{34.6} 
  & \textbf{99.7} & 8.3 & \textbf{74.9} & 18.3 & \textbf{50.3} & \textbf{42.5} \\

  & $\mathcal{B}^4$(4,3) 
  & \textbf{24.1} & 79.9 & 53.3 & \textbf{30.6} & \textbf{5.0} & \textbf{14.7} & \textbf{34.6}
  & 99.4 & \textbf{9.9} & 72.8 & 18.0 & 50.0 & 42.3 \\

  & $\mathcal{B}^4$(5,3) 
  & 22.9 & 78.6 & \textbf{61.7} & 22.2 & 3.3 & 12.3 & 33.5 
  & 98.8 & 9.6 & 72.5 & \textbf{19.6} & 50.1 & 41.8 \\

\midrule

\multicolumn{15}{@{}l}{\cellcolor{gray!8}\textit{Qwen2.5-Coder-7B-Instruct}} \\[2pt]

Base Model & -- 
  & 15.1 & 83.6 & 60.0 & 23.3 & \textbf{18.3} & 8.0 & 34.7 
  & 84.4 & 6.2 & 69.1 & 14.6 & 43.6 & 39.1 \\

\quad+Offline & -- 
  & 15.1 & 82.8 & 51.7 & 21.7 & 5.0 & 10.3 & 31.1 
  & 87.0 & 7.7 & 69.3 & 14.8 & 44.7 & 37.9 \\

\quad+Online & -- 
  & 18.1 & 81.2 & 61.7 & 18.9 & 8.3 & 9.3 & 32.9 
  & 88.9 & 7.4 & 68.5 & 14.6 & 44.8 & 38.9 \\

\quad+ZeroCoder 
  & MaxPass 
  & 14.5 & 84.1 & 66.7 & 23.3 & 13.3 & 8.7 & 35.1 
  & 74.1 & 18.9 & 72.0 & \textbf{37.3} & 50.6 & 42.8 \\

  & CodeT 
  & \textbf{20.5} & \textbf{85.4} & 68.3 & 21.1 & 11.7 & 11.0 & 36.3 
  & \textbf{91.3} & 15.2 & \textbf{72.5} & 36.8 & \textbf{53.9} & \textbf{45.1} \\

  & $\mathcal{B}^4$(4,3) 
  & \textbf{20.5} & 84.9 & \textbf{70.0} & \textbf{31.1} & \textbf{18.3} & \textbf{13.7} & \textbf{39.8} 
  & 75.2 & \textbf{20.1} & 69.6 & 36.5 & 50.3 & 45.0 \\

  & $\mathcal{B}^4$(5,3) 
  & \textbf{20.5} & 83.6 & 66.7 & 21.1 & 16.7 & 9.3 & 36.3 
  & 68.4 & \textbf{20.1} & 70.9 & 35.7 & 48.8 & 42.5 \\

\bottomrule
\end{tabular}%
}
\end{table*}

Table~\ref{tab:label-free} summarizes the performance of \app and other baselines. 

First, \app with static selectors is already effective in a fully label-free setting. For example, on Qwen2.5-Coder-7B-Instruct, it improves code generation and test generation over the base model by up to 14.5\% and 23.9\%, respectively. These results suggest that self-generated code-test interactions can produce useful supervision, leading to substantial improvements in both roles.

To further figure out the reasons for \app’s better performance, we manually inspect the code and test outputs generated by the base model (Qwen2.5-Coder-7B-Instruct) and its \app-trained variant (using CodeT as the selector). Based on our inspection, we find that \app has two major advantages:
\begin{figure}[t]
    \centering
    \includegraphics[width=\columnwidth]{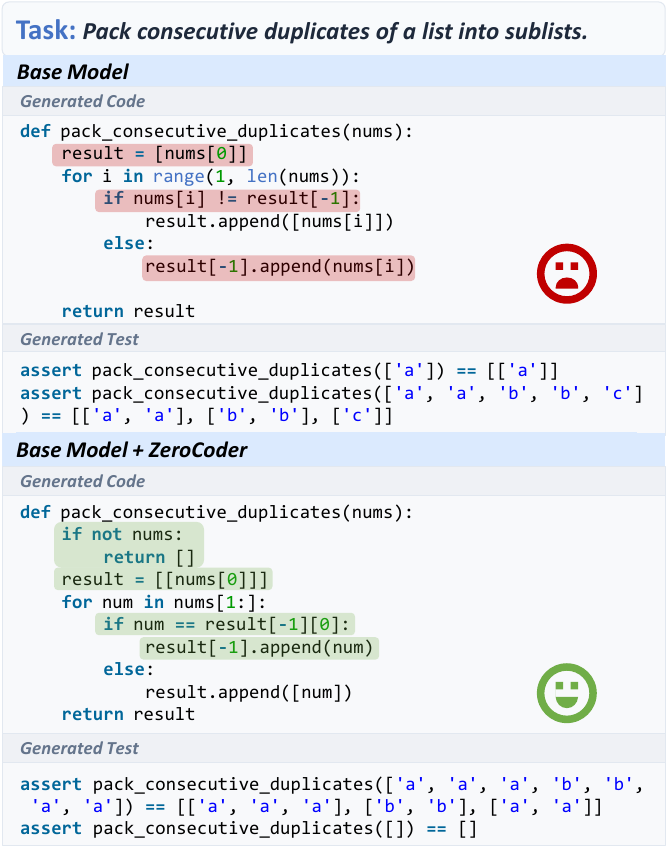}
    \caption{A case study demonstrating the advantages of \app.}
    \label{fig:rq1-case}
\end{figure}

\textit{\app can suppress invalid generations through external execution feedback.}
In \app, every candidate solution is executed against sampled tests, so structurally flawed code receives negative feedback through low rewards.
As shown in Figure~\ref{fig:rq1-case}, the task requires grouping consecutive duplicate elements of a list into sublists.  The base model produces an invalid implementation by initializing \texttt{result} as a flat list, which propagates into subsequent errors: the comparison checks against a raw element, and the \texttt{append} call fails.
In contrast, under \app, such faulty code is exposed during execution against sampled tests, receives low rewards, and is therefore suppressed during training.

\textit{The co-evolved tester generates increasingly discriminative tests that, in turn, push the coder toward more robust solutions.}
In the same example, the base model generates weak tests that cover simple functionality, such as a single-element input and a standard multi-group list, but miss boundary conditions.
After training, the tester generates a boundary test for the empty-list input together with a longer and more challenging grouped sequence.
These discriminative tests not only validate basic behavior, but also distinguish correct solutions from plausible ones that fail on edge cases. As a result, the coder is pushed to satisfy stricter behavioral requirements, leading to more robust code generation.

Second, relying solely on offline or online test-driven RL without co-evolving the tester degrades code-generation performance. For example, on Qwen2.5-Coder-7B-Instruct, the offline and online RL baselines reduce code generation performance by 10.4\% and 5.2\% relative to the base model, respectively. A plausible explanation is that, in both baselines, the tester is fixed rather than improved alongside the increasingly capable coder. As training progresses, such initially weak tests become insufficient to distinguish truly correct solutions from plausible but faulty ones, leading to noisy or low-information reward signals that ultimately hinder learning.

Lastly, no single static selector consistently dominates across all evaluated benchmarks. For example, on Qwen2.5-Coder-7B-Instruct, \app with $\mathcal{B}^4(4,3)$ achieves the best coding performance, improving over the base model by 14.5\% on average, whereas CodeT yields the best test-generation performance, outperforming the base model by 23.9\%. We hypothesize that this is because their fixed inductive biases cannot stay calibrated to the evolving reliability of the coder and tester during training, leading to unstable learning signals.

\begin{figure}
    \centering
    \vspace{-2mm}
    \includegraphics[width=0.9\linewidth]{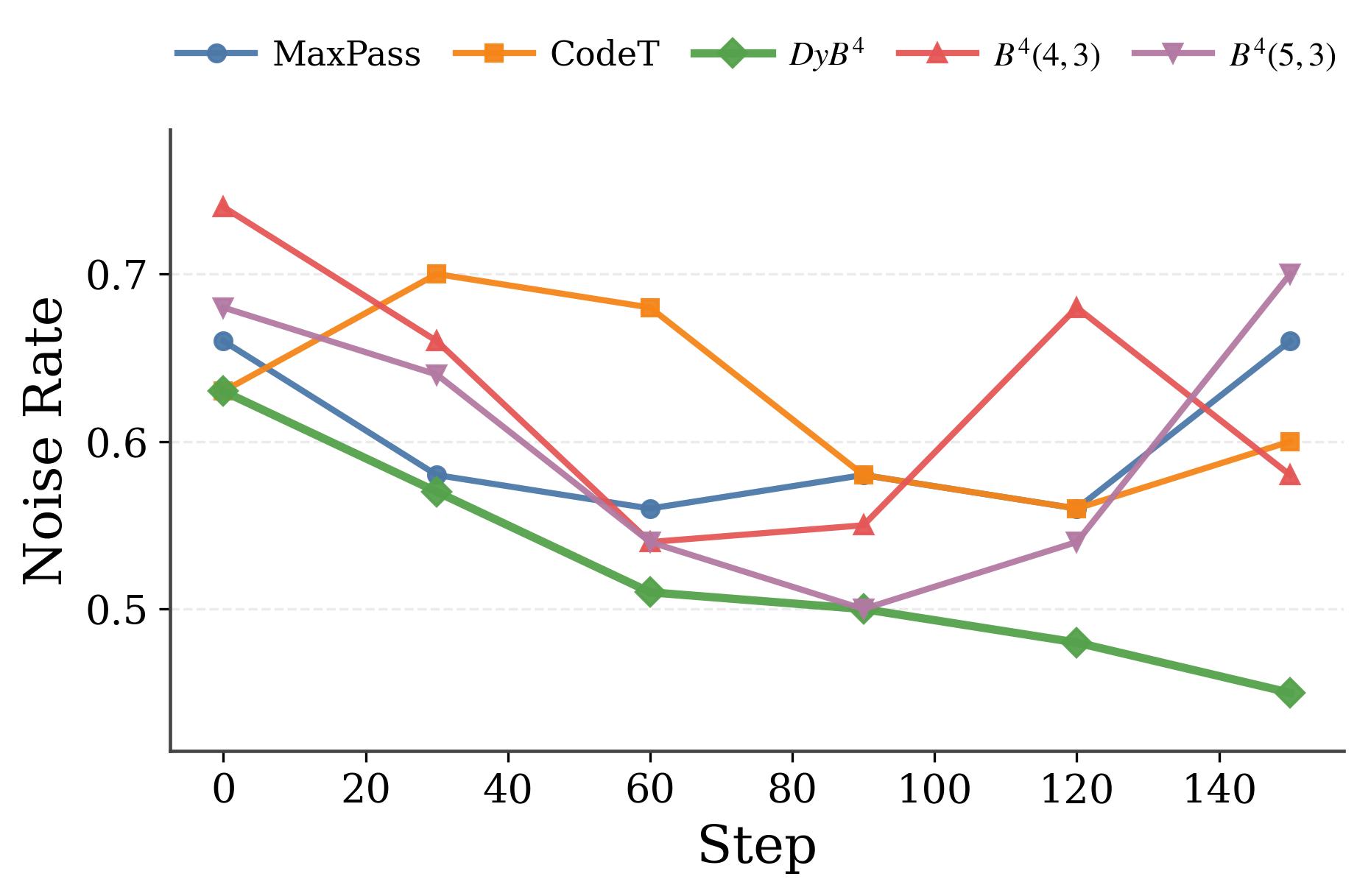}
    \vspace{-4mm}
    \caption{Analysis of the selection noise rate of static selectors and \appdy over training steps.}
    \label{fig:static_selector_drift}
\end{figure}

To make this explicit, we track the selection noise rate of each selector over training.
We randomly sample 300 training instances and, at each checkpoint, measure selection noise rate as follows: $\text{noise}_k=1 - Prec_k(\mathrm{Sel})$, where $Prec$ is the fraction of selected solutions that pass the official test suite.
Figure~\ref{fig:static_selector_drift} shows that $\text{noise}_k$ fluctuates and can even increase over training for static selectors, indicating that static selectors may become miscalibrated and even yield degraded selection quality. 
We define this phenomenon as \textit{selector drift}, i.e., the progressive degradation of selection accuracy due to fixed prior assumptions that fail to align with the evolving distribution of candidate solutions.
In contrast, \appdy continuously adapts to these shifts, maintaining a stable precision gap throughout the co-evolutionary process.

\begin{findingbox}
\textbf{Answer to RQ1:} In the fully label-free setting, \app with static selectors consistently improves both code and test generation, achieving average relative improvements of 8.6\% and 58.2\% over the base models across all three models, respectively. This demonstrates that self-generated code-test interactions can serve as effective supervision. In contrast, offline and online test-driven RL without co-evolving the tester degrades code-generation performance. Analysis of selection noise rates further reveals that static selectors suffer from progressive miscalibration during co-evolution, motivating dynamic calibration.
\end{findingbox}

\subsection{RQ2: Dynamic Calibration with \appdy}
\label{sec:appdy}

\begin{table*}[t]
\footnotesize
\centering
\caption{Performance of ZeroCoder$^\dagger$ with $\mathit{Dy}\mathcal{B}^4$ and static selectors under a \textbf{matched small-label} setting ($|\mathcal{H}|=10$) across three base models. All ZeroCoder$^\dagger$ variants share the same labeled set $\mathcal{H}$: static selectors use it only for supervised reward, while $\mathit{Dy}\mathcal{B}^4$ additionally uses it for Bayesian prior calibration. Performance differences from the label-free variants are due to this additional supervision.}
\label{tab:label-assisted}
\resizebox{0.9\textwidth}{!}{%
\begin{tabular}{@{}ll ccccccc cccc c c@{}}
\toprule
\multirow{3}{*}{\textbf{Model}} & \multirow{3}{*}{\textbf{Selector}} 
& \multicolumn{7}{c}{\textbf{Code Generation}} 
& \multicolumn{4}{c}{\textbf{Test Generation}} 
& \multirow{3}{*}{\textbf{TAvg}} 
& \multirow{3}{*}{\textbf{Avg}} \\
\cmidrule(lr){3-9} \cmidrule(lr){10-13}
& & \multirow{2}{*}{LCB} & \multirow{2}{*}{MBPP} & \multicolumn{3}{c}{Apps} & \multirow{2}{*}{CF.} & \multirow{2}{*}{CAvg} 
& \multicolumn{2}{c}{LCB} & \multicolumn{2}{c}{MBPP} & & \\
\cmidrule(lr){5-7} \cmidrule(lr){10-11} \cmidrule(lr){12-13}
& & & & Intro. & Inter. & Comp. & & & ACC. & Mut. & ACC. & Mut. & & \\
\midrule

\multicolumn{15}{@{}l}{\cellcolor{gray!8}\textit{Qwen2.5-1.5B-Instruct}} \\[2pt]

Base Model & -- 
  & 4.8 & 60.3 & 31.7 & 6.1 & 3.3 & 1.7 & 18.0 
  & 17.5 & 3.0 & 42.7 & 7.2 & 17.6 & 17.8 \\

\quad+ZeroCoder$^\dagger$ 
  & MaxPass 
  & 4.8 & \textbf{66.7} & 33.3 & 7.2 & 3.3 &  \textbf{2.7} & 19.7 
  & 62.5 & 17.3 & 66.1 & 28.8 & 43.7 & 31.7 \\

  & CodeT 
  & 4.8 & 64.6 & 35.0 & 7.2 & 3.3 & 2.3 & 19.5 
  & 64.7 & 15.2 &  \textbf{68.5} & 30.4 & 44.7 & 32.1 \\

  & $\mathcal{B}^4$(4,3) 
  &  \textbf{5.4} & 63.5 & 31.7 & 6.7 & 3.3 & 2.0 & 18.8 
  & \textbf{69.3} & 13.6 & 67.2 & 24.6 & 43.7 & 31.2 \\

  & $\mathcal{B}^4$(5,3) 
  &  \textbf{5.4} & 62.2 & 33.3 & 6.1 & \textbf{5.0} & 2.0 & 19.0 
  & 67.8 & 14.6 & 66.7 & 25.9 & 43.8 & 31.4 \\

  \rowcolor{blue!5}
  & \appdy 
  &  \textbf{5.4} & 65.6 & \textbf{40.0} & \textbf{10.0} & 3.3 & 2.0 & \textbf{21.1} 
  & 67.5 & \textbf{17.6} & 68.0 & \textbf{30.7} & \textbf{46.0} & \textbf{33.5} \\

\midrule

\multicolumn{15}{@{}l}{\cellcolor{gray!8}\textit{Qwen3-4B}} \\[2pt]

Base Model & -- 
  & 21.7 & 79.9 & 50.0 & 21.1 & 1.7 & 8.3 & 30.4 
  & 99.4 & 9.3 & 72.4 & 18.0 & 49.8 & 40.1 \\

\quad+ZeroCoder$^\dagger$ 
  & MaxPass 
  & 24.1 &  \textbf{80.2} & 55.0 & 22.2 & 1.7 & 12.7 & 32.6 
  & 99.7 & \textbf{9.6} & 72.5 & 18.0 & 49.9 & 41.3 \\

  & CodeT 
  & 24.7 & 78.6 & 56.7 & 27.8 &  \textbf{3.3} & 11.7 & 33.8 
  & 99.7 & \textbf{9.6} & \textbf{75.7} & 16.7 & 50.4 & 42.1 \\

  & $\mathcal{B}^4$(4,3) 
  & 22.9 & 79.1 &  \textbf{58.3} & 27.2 & 1.7 & 10.7 & 33.3 
  & 99.7 & \textbf{9.6} & 74.3 & 16.4 & 50.0 & 41.7 \\

  & $\mathcal{B}^4$(5,3) 
  & 24.7 & 77.8 & 50.0 & 26.7 &  \textbf{3.3} & 12.3 & 32.5 
  & 99.7 & 8.7 & 75.9 & 16.4 & 50.2 & 41.3 \\

  \rowcolor{blue!5}
  & \appdy 
  & \textbf{25.3} & 79.9 & 56.7 & \textbf{35.6} & \textbf{3.3} & \textbf{14.3} & \textbf{35.8} 
  & \textbf{100.0} & \textbf{9.6} & 74.6 & \textbf{20.9} & \textbf{51.3} & \textbf{43.6} \\

\midrule

\multicolumn{15}{@{}l}{\cellcolor{gray!8}\textit{Qwen2.5-Coder-7B-Instruct}} \\[2pt]

Base Model & -- 
  & 15.1 & 83.6 & 60.0 & 23.3 & 18.3 & 8.0 & 34.7 
  & 84.4 & 6.2 & 69.1 & 14.6 & 43.6 & 39.1 \\

\quad+ZeroCoder$^\dagger$ 
  & MaxPass 
  & 18.7 & 83.3 & 65.0 & 25.6 & 13.3 & 10.7 & 36.1 
  & 87.3 & 15.2 & 73.8 & 37.6 & 53.5 & 44.8 \\

  & CodeT 
  & 19.3 & 83.6 & 65.0 & 27.8 & 15.0 & \textbf{13.7} & 37.4 
  & \textbf{88.6} & 14.7 & \textbf{76.5} & 27.5 & 51.8 & 44.6 \\

  & $\mathcal{B}^4$(4,3) 
  & 18.7 & \textbf{85.4} & 66.7 & 25.0 & 15.0 & 12.0 & 37.1 
  & 77.4 & \textbf{19.5} & 70.9 & 36.5 & 48.6 & 42.9 \\

  & $\mathcal{B}^4$(5,3) 
  & 18.1 & 83.3 & 60.0 & 22.8 & 16.7 & 9.3 & 35.0 
  & 82.7 & 18.0 & 72.0 & 36.8 & 52.4 & 43.7 \\

  \rowcolor{blue!5}
  & \appdy 
  & \textbf{23.5} & 84.9 & \textbf{75.0} & \textbf{32.8} & \textbf{23.3} & \textbf{13.7} & \textbf{42.2} 
  & 85.8 & 18.0 & 74.1 & \textbf{38.6} & \textbf{54.1} & \textbf{48.2} \\

\bottomrule
\end{tabular}%
}
\end{table*}

We answer RQ2 by evaluating whether dynamic calibration further improves \app under a matched small-label setting. Specifically, we equip all selectors with the same small labeled set $\mathcal{H}$ ($|\mathcal{H}|=10$), so that performance differences cannot be attributed to unequal supervision budgets. For static selectors, $\mathcal{H}$ is used only to provide supervised rewards during training. For \appdy, the same set is used both for supervised rewards and for dynamically recalibrating the selector. On $\mathcal{H}$, rewards are computed from ground-truth supervision: for the coder, a solution receives reward 1 iff it passes all ground-truth tests, and 0 otherwise; for the tester, the reward follows \S\ref{sec:coevo}, except that the consensus solution $s^\star$ is replaced by the ground-truth solution.

Under this matched supervision budget, \appdy further brings robust and consistent improvements in both code generation and test generation across all three models. As shown in Table~\ref{tab:label-assisted}, on Qwen2.5-Coder-7B-Instruct, \app with \appdy achieves average relative improvements of 21.6\% and 24.3\% over the base model in code generation and test generation, respectively. Notably, \appdy consistently outperforms all static selectors under the same labeled budget, indicating that its gains come from better-calibrated reward construction rather than simply from access to labeled data.

A key reason is that \appdy continually adapts the selector to the evolving reliability of the coder and tester during training. As shown in Figure~\ref{fig:static_selector_drift}, while static selectors can fluctuate in selection quality and even deteriorate over time, \appdy maintains a consistently lower noise rate and a sustained selection-quality advantage throughout co-evolution. 

\begin{findingbox}
\textbf{Answer to RQ2:} Under a matched small-label budget ($|\mathcal{H}|=10$), \appdy consistently outperforms all static selectors across all three models, achieving average relative improvements of 18.8\% and 62.7\% over the base models in code and test generation, respectively. These gains arise because dynamic calibration keeps the selector aligned with the evolving reliability of the coder and tester throughout training.
\end{findingbox}

\subsection{RQ3: Component, Sensitivity, and Efficiency Analysis}
\label{sec:rq3_analysis}
We answer RQ3 by analyzing the contribution of each component through ablation studies and sensitivity analysis, profiling training efficiency, and comparing against oracle-supervised training. Due to computational constraints, we conduct analytical experiments on Qwen2.5-coder-7B-Instruct, the strongest-performing model in our evaluation, unless explicitly stated otherwise. 

\paragraph{Effect of the Mutation-Based Reward}
\label{sec:abl}
To analyze whether gains come merely from improving test validity, or from encouraging tests to be discriminative enough to provide useful feedback to the coder, we ablate the mutation-based reward component in the tester objective.
Concretely, we remove the mutation-based term $r^{\text{mut}}$ from \app+\appdy and re-train all models with the other settings unchanged.
The results are summarized in Table~\ref{tab:ablation}. 

Generating discriminative tests is essential for the continuous co-evolution of the coder and tester. As shown in the table, removing the mutation reward leads to a degradation in code generation performance. For example, the average coding performance of Qwen2.5-Coder-7B-Instruct drops from 42.2\% to 35.3\% when the mutation reward is removed. This may be because, without mutation, the tester fails to explore discriminative cases, leading to trivial tests that provide weak training signals to the coder.

\begin{table}[t]
\small
\centering
\caption{Ablation study and oracle comparison across three base models. ``w/o $r^{\text{mut}}$'' removes the mutation-based discriminativeness reward from the Tester objective. ``Oracle'' trains with ground-truth solutions and tests.}
\vspace{-2mm}
\label{tab:ablation}
\setlength{\tabcolsep}{6pt}
\renewcommand{\arraystretch}{1.05}
\resizebox{0.85\columnwidth}{!}{%
\begin{tabular}{l ccc}
\toprule
\textbf{Model} & \textbf{CAvg} & \textbf{TAvg} & \textbf{Avg} \\
\midrule
\multicolumn{4}{l}{\cellcolor{gray!8}\textit{Qwen2.5-1.5B}} \\[2pt]
Base Model                                          & 18.0          & 17.6          & 17.8 \\
\quad+ZeroCoder (\appdy) w/o $r^{\text{mut}}$            & 19.2          & 44.8          & 32.0 \\
\rowcolor{blue!5}
\quad+ZeroCoder (\appdy)                                 & \underline{21.1} & \underline{46.0} & \underline{33.5} \\
\arrayrulecolor{gray!40}\specialrule{0.4pt}{2pt}{2pt}\arrayrulecolor{black}
\quad\textcolor{gray!90}{+Oracle (w/ GT)}                 & \textcolor{gray!90}{\textit{\textbf{21.6}}} & \textcolor{gray!90}{\textit{\textbf{47.1}}} & \textcolor{gray!90}{\textit{\textbf{34.3}}} \\
\midrule
\multicolumn{4}{l}{\cellcolor{gray!8}\textit{Qwen3-4B}} \\[2pt]
Base Model                                          & 30.4          & 49.8          & 40.1 \\
\quad+ZeroCoder (\appdy) w/o $r^{\text{mut}}$            & 33.3          & 45.8          & 39.6 \\
\rowcolor{blue!5}
\quad+ZeroCoder (\appdy)                                 & \textbf{35.8} & \textbf{51.3} & \textbf{43.6} \\
\arrayrulecolor{gray!40}\specialrule{0.4pt}{2pt}{2pt}\arrayrulecolor{black}
\quad\textcolor{gray!90}{+Oracle (w/ GT)}                 & \textcolor{gray!90}{\textit{\underline{35.4}}} & \textcolor{gray!90}{\textit{\underline{50.8}}} & \textcolor{gray!90}{\textit{\underline{43.1}}} \\
\midrule
\multicolumn{4}{l}{\cellcolor{gray!8}\textit{Qwen2.5-Coder-7B-Instruct}} \\[2pt]
Base Model                                          & 34.7          & 43.6          & 39.1 \\
\quad+ZeroCoder (\appdy) w/o $r^{\text{mut}}$            & 35.3          & 51.9          & 43.6 \\
\rowcolor{blue!5}
\quad+ZeroCoder (\appdy)                                 & \underline{42.2} & \underline{54.1} & \underline{48.2} \\
\arrayrulecolor{gray!40}\specialrule{0.4pt}{2pt}{2pt}\arrayrulecolor{black}
\quad\textcolor{gray!90}{+Oracle (w/ GT)}                 & \textcolor{gray!90}{\textit{\textbf{43.2}}} & \textcolor{gray!90}{\textit{\textbf{55.0}}} & \textcolor{gray!90}{\textit{\textbf{49.1}}} \\
\bottomrule
\end{tabular}%
}\\[4pt]
{\small \textbf{Bold}: best result; \underline{Underline}: second best.}
\end{table}

\paragraph{Impact of Interaction Diversity}

We explore how interaction diversity in the training set affects model performance.
Specifically, within \app+\appdy, we filter the training data using different rank thresholds $\tau \in {0,1,2,4}$ and train models with the same configuration as in the main experiments. Notably, $\tau=0$ denotes training on the original dataset.

\begin{figure}[t]
\vspace{-1.2em}
    \centering
    \begin{subfigure}[t]{0.49\linewidth}
        \centering
        \includegraphics[width=\linewidth]{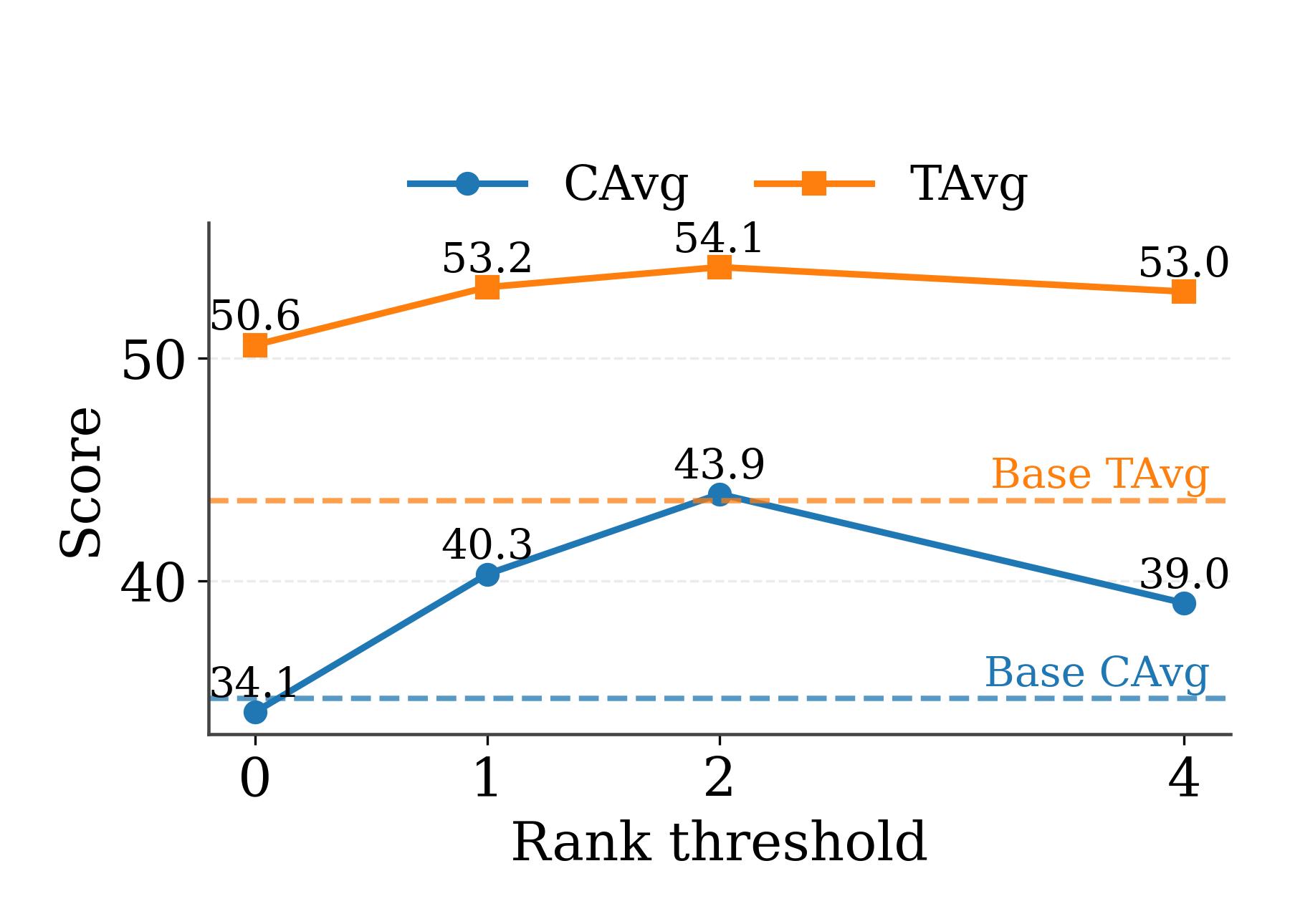}
        \caption{Interaction diversity threshold $\tau$.}
        \label{fig:diversity}
    \end{subfigure}
    \begin{subfigure}[t]{0.49\linewidth}
        \centering
        \includegraphics[width=\linewidth]{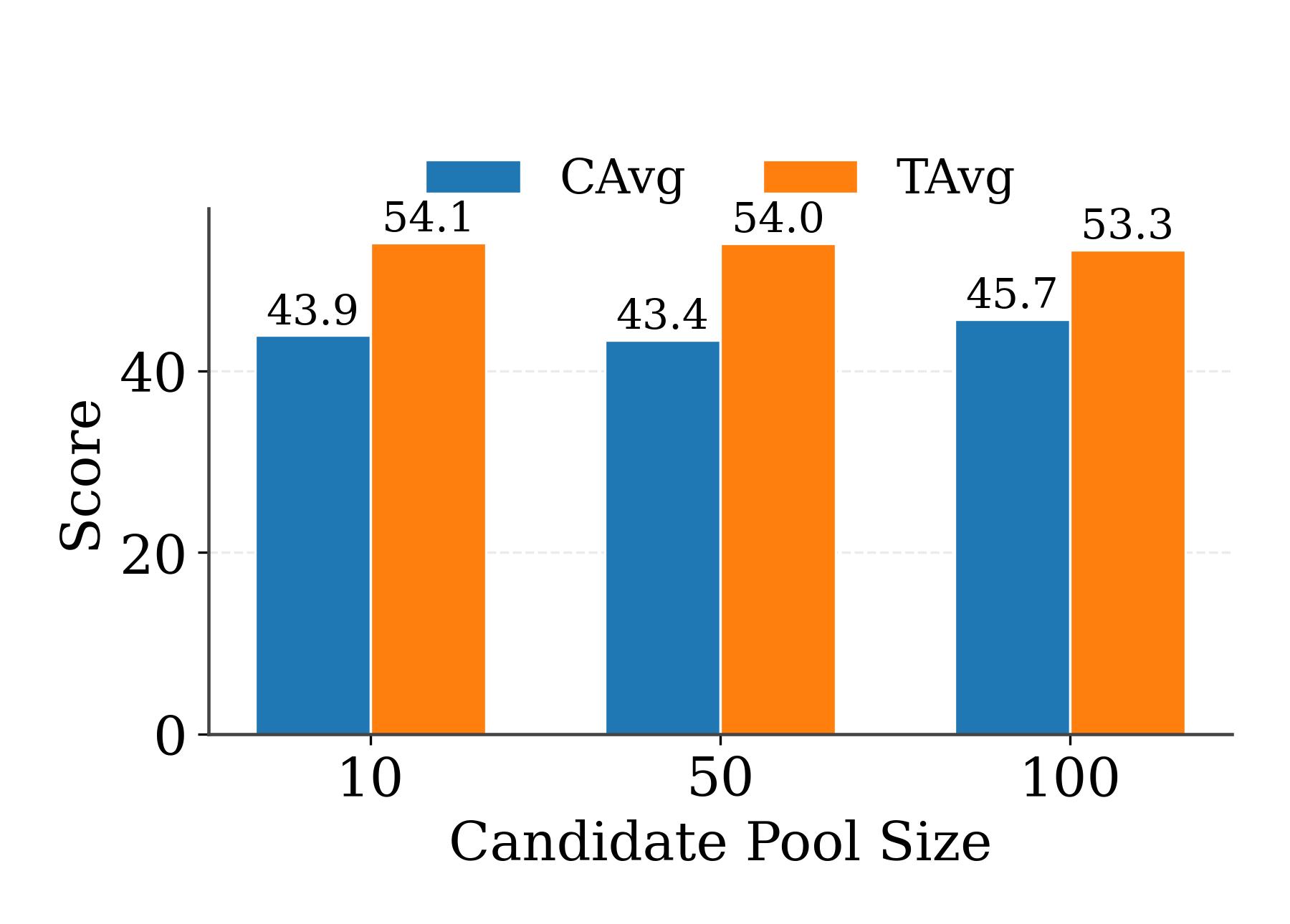}
        \caption{Calibration set size $|\mathcal{H}|$.}
        \label{fig:sensitivity}
    \end{subfigure}
    \vspace{-2mm}
    \caption{Sensitivity analysis of \app: (a) performance under varying interaction-diversity filtering thresholds $\tau$; (b) under different calibration set sizes $|\mathcal{H}|$}
    \label{fig:analysis_subfigures}
\end{figure}

As shown in Figure~\ref{fig:analysis_subfigures}(\subref{fig:diversity}), both code generation and test generation performance first improve and then degrade as the threshold becomes stricter.
Notably, the performance consistently remains superior to that of the base model and the model trained on the unfiltered dataset, suggesting the effectiveness of the filtering method.
The performance degradation at high thresholds is likely attributed to data scarcity.
A strict filter reduces the volume of available training data, causing the model to overfit.
To substantiate this hypothesis, we quantify the number of remaining training instances under each threshold in Table~\ref{tab:filter_stats}, where we observe a sharp drop in data volume at $\tau=4$.

\paragraph{Sensitivity to Calibration Set Size}
\label{sec:abl_cal}

We examine the sensitivity of \appdy to the size of the calibration set $\mathcal{H}$ by varying $|\mathcal{H}| \in \{10,50,100\}$. As depicted in Figure~\ref{fig:analysis_subfigures}(\subref{fig:sensitivity}), \appdy is relatively insensitive to $|\mathcal{H}|$: even 10 labeled instances achieve performance comparable to substantially larger calibration sets. Although such a small set provides a weaker calibration signal than larger alternatives, it is already sufficient in practice for effective recalibration. This observation is consistent with prior findings on $\mathcal{B}^4$~\cite{chen2024b4} that good performance can often be achieved without highly precise hyperparameter tuning. Therefore, a small calibration set is sufficient for \appdy to identify a well-performing configuration at each recalibration step.

\paragraph{Training Efficiency Analysis.}
To understand the computational overhead, we profile the per-step wall-clock time of \app with $\appdy$. We decompose the total time into five categories: (i)~\textit{Rollout}, the time for generating candidate solutions and tests; (ii)~\textit{Execution}, the time for executing all solution-test pairs to construct the passing matrix; (iii)~\textit{Mutation}, the time for generating and executing mutants against candidate tests (on average, 13.4 mutants are generated per sample); (iv)~\textit{$\appdy$ Calibration}, the time for recalibrating the Bayesian prior at each step; and (v)~\textit{RL Training}, the time for computing advantages, the backward pass, and optimizer updates. The breakdown is summarized in Table~\ref{tab:efficiency}.

\begin{table}[t]
\centering
\caption{Number of retained APPS training problems after rank-based filtering under different thresholds $\tau$.}
\label{tab:filter_stats}
\small
\begin{tabular}{lrrrr}
\toprule
\textbf{Model} & $\tau=0$ & $\tau=1$ & $\tau=2$ & $\tau=4$ \\
\midrule
Qwen2.5-1.5B-Instruct      & 5000 & 2283 & 1377 & 217 \\
Qwen3-4B                    & 5000 & 2546 & 990  & 279 \\
Qwen2.5-Coder-7B-Instruct  & 5000 & 2615 & 1396 & 428 \\
\bottomrule
\end{tabular}
\end{table}

\begin{table}[t]
\centering
\caption{Per-step wall-clock time breakdown of \app with $\appdy$ on Qwen2.5-Coder-7B-Instruct.}
\small
\label{tab:efficiency}
\begin{tabular}{lrr}
\toprule
\textbf{Component} & \textbf{Time (s)} & \textbf{Fraction (\%)} \\
\midrule
Rollout              & 32.1  & 5.6  \\
Execution            & 273.0 & 47.6 \\
Mutation             & 204.7 & 35.7 \\
$\appdy$ Calibration & 11.4  & 2.0  \\
RL Training          & 52.8  & 9.2  \\
\midrule
\textbf{Total}       & \textbf{574.0} & \textbf{100.00} \\
\bottomrule
\end{tabular}

\end{table}

As shown in Table~\ref{tab:efficiency}, most of the runtime is dominated by verifier-side execution and mutation, while dynamic calibration contributes only a small fraction. In particular, $\appdy$ calibration accounts for only 2.0\% of the per-step time. Most of the wall-clock time is spent on Execution (47.6\%) and Mutation (35.7\%), which are both verifier-side program-analysis workloads. This is expected in our setting, since \app improves supervision quality through richer code-test interaction signals. Importantly, these components are embarrassingly parallel across problems and scale naturally with the number of execution workers, making them straightforward to accelerate with additional compute.

\paragraph{Comparison with the Oracle-Supervised Reference}
We compare \app against an oracle-supervised reference variant that has access to ground-truth solutions and tests from the dataset (\textit{Oracle}). The reward computation follows the same protocol as in RQ2 when $\mathcal{H}$ is used to provide supervised training rewards. The performance comparison is detailed in Table~\ref{tab:ablation}.
We observe that \app achieves performance comparable to the oracle-supervised models. For example, for Qwen2.5-coder-7B-Instruct, \app attains an average coder performance of 42.2\%, comparable to 43.2\% under the oracle setting. This result suggests that the synthetic interaction data produced by our co-evolutionary framework enables effective learning without relying on ground-truth supervision.

\begin{findingbox}
\textbf{Answer to RQ3:} \app’s gains are driven by two key design choices: rank-based pre-filtering, which improves reward informativeness by removing low-information problems, and the mutation-based tester reward, which promotes discriminative tests and thereby strengthens coder learning. Moreover, \appdy is label-efficient, and its final performance is competitive with that of oracle-supervised training.
\end{findingbox}

\section{Related Work}
\textit{Self-Rewarding in Large Language Models.}
Self-rewarding methods improve LLMs without external labels by deriving training signals from the model’s own outputs, and can be broadly grouped into three categories.
The first category constructs rewards from model-side self-assessment signals. For example, methods based on confidence~\citep{prabhudesai2025maximizing,jang2025self,zhang2025right} use the model's own uncertainty estimates as a proxy for correctness, rewarding outputs that the model is more confident about. 
The second category derives supervision from consensus among multiple candidates, such as majority voting~\citep{zuo2025ttrl,huang2025r}, treating cross-sample agreement as a proxy for quality. Both categories rely on internal model agreement, which may reinforce incorrect answers and induce performance collapse~\citep{shafayat2025can}.
The third category, relevant to code generation, grounds rewards in program execution. Rather than relying on internal agreement, these methods generate tests and evaluate candidate solutions by running them~\cite{zeng2025acecoder,zhang2025codedpo,liu2024dstc,gorinski2023automatic}. 
For example, AceCoder~\citep{zeng2025acecoder} synthesizes test cases via a separate pipeline and uses their pass rates to construct binary rewards for RL training of the coder. CodeDPO~\citep{zhang2025codedpo} uses self-generated tests with a PageRank-inspired algorithm to construct preference pairs for coder optimization. 
A common limitation is that the test generator remains fixed, so as the coder improves, the static tests lose discriminative power and training signal diminishes. In contrast, \app co-evolves both roles so that test quality scales with coder capability.

\textit{Reranking and Selection for Solutions.}
Selecting the optimal solution from a pool of generated candidates is crucial for code generation.
Existing approaches can be broadly categorized into execution-based and non-execution-based methods.
Execution-based methods generate test cases, execute candidate programs, and select solutions based on pass statistics~\citep{li2022competition,shi2022natural,chencodet,chen2024b4}.
Non-execution-based methods train neural rerankers to estimate correctness without running the code~\citep{inala2022fault,zhang2023coder}.
More recently, Bayesian selection strategies, i.e., $\mathcal{B}^4$, formalize solution selection when both solutions and tests are plausible, grounding selection in posterior reasoning over the observed passing matrix~\citep{chen2024b4}.
However, prior work primarily treats solution selection as an inference-time technique, and training-time selection is often instantiated with simple heuristics such as majority voting to form pseudo-labels, overlooking how selector quality directly affects the fidelity of the resulting supervision~\citep{zuo2025ttrl,huang2025r}.
In contrast, \app integrates selection into the training loop and uses the resulting consensus structure to construct rewards.

\textit{Reinforcement Learning for Language Models.}
RLVR has emerged as a widely adopted paradigm for improving LLM reasoning in code generation~\citep{guo2025deepseek,yang2025qwen3,abdin2025phi}, typically by optimizing policies against objective outcome-based signals~\citep{shao2024deepseekmath,yu2025dapo,zheng2025group}.
However, most RLVR pipelines for code still rely on trusted verifiers to provide reliable rewards.
As these supervised data sources saturate, further improvements in LLMs become difficult. To address this, \app extends RLVR into a co-evolutionary framework without requiring ground-truth solutions or tests.

\section{Threats to Validity}
\textit{Threats to external validity} concern the generalizability of our findings. To improve coverage, we evaluate \app across three LLMs from different families and scales, including Qwen2.5-1.5B-Instruct, Qwen3-4B, and Qwen2.5-Coder-7B-Instruct, and consider both the fully label-free setting and the small-label setting. In addition, key ablations, such as removing the mutation-based reward and comparing against oracle-supervised training, are validated across all three models. Nevertheless, our experiments are limited to models of up to 7B parameters due to computational constraints. In future work, we plan to explore a broader range of model series to further validate the generalizability of our approach.

\textit{Threats to internal validity} relate to the prompt templates used to instantiate the Coder and Tester, which may not be optimal. We design these prompts following prior work~\citep{jain2025livecodebench} and keep them fixed across all compared methods to ensure fairness. While more advanced prompting strategies, such as chain-of-thought, may further improve performance, they are orthogonal to the core contribution of our framework and could be incorporated in future extensions.

\textit{Threats to construct validity} concern whether our measurements faithfully capture the underlying concepts of interest. In \app, we use the rank of the passing matrix as a proxy for interaction diversity and reward informativeness. Although this proxy is empirically effective, matrix rank is only an approximation and may not fully characterize all aspects of supervision quality. For example, two interaction matrices with the same rank may still differ substantially in the usefulness of the rewards they induce. 
Future work could explore more expressive characterizations of interaction structure, such as covariance-based statistics~\cite{lam2020high} and spectral properties~\cite{roy2007effective} of the passing matrix.

\section{Conclusion}
We present \app, a label-free co-evolutionary framework that improves code generation and test generation from self-generated code-test interactions. By converting passing matrices into role-specific rewards, \app enables the coder and tester to improve without relying on ground-truth supervision. Results show that, when instantiated with static selectors, \app already delivers consistent gains, demonstrating that code-test interaction alone can serve as an effective training signal.
We also identify selector drift as a key source of reward noise in non-stationary co-evolution. To address this, we introduce \appdy, a dynamically calibrated Bayesian selector that uses a small labeled set for recalibration. Across three models and six benchmarks, \appdy further improves over static-selector variants and achieves the strongest overall performance. Together, these results show that label-free co-evolution is effective on its own, while lightweight calibration can further strengthen it. 

\section*{Data Availability}
Our code is available: https://doi.org/10.5281/zenodo.19247497.

\bibliographystyle{ACM-Reference-Format}
\bibliography{custom}


\begin{thebibliography}{53}


\ifx \showCODEN    \undefined \def \showCODEN     #1{\unskip}     \fi
\ifx \showISBNx    \undefined \def \showISBNx     #1{\unskip}     \fi
\ifx \showISBNxiii \undefined \def \showISBNxiii  #1{\unskip}     \fi
\ifx \showISSN     \undefined \def \showISSN      #1{\unskip}     \fi
\ifx \showLCCN     \undefined \def \showLCCN      #1{\unskip}     \fi
\ifx \shownote     \undefined \def \shownote      #1{#1}          \fi
\ifx \showarticletitle \undefined \def \showarticletitle #1{#1}   \fi
\ifx \showURL      \undefined \def \showURL       {\relax}        \fi
\providecommand\bibfield[2]{#2}
\providecommand\bibinfo[2]{#2}
\providecommand\natexlab[1]{#1}
\providecommand\showeprint[2][]{arXiv:#2}

\bibitem[Abdin et~al\mbox{.}(2025)]%
        {abdin2025phi}
\bibfield{author}{\bibinfo{person}{Marah Abdin}, \bibinfo{person}{Sahaj Agarwal}, \bibinfo{person}{Ahmed Awadallah}, \bibinfo{person}{Vidhisha Balachandran}, \bibinfo{person}{Harkirat Behl}, \bibinfo{person}{Lingjiao Chen}, \bibinfo{person}{Gustavo de Rosa}, \bibinfo{person}{Suriya Gunasekar}, \bibinfo{person}{Mojan Javaheripi}, \bibinfo{person}{Neel Joshi}, {et~al\mbox{.}}} \bibinfo{year}{2025}\natexlab{}.
\newblock \showarticletitle{Phi-4-reasoning technical report}.
\newblock \bibinfo{journal}{\emph{arXiv preprint arXiv:2504.21318}} (\bibinfo{year}{2025}).
\newblock


\bibitem[Austin et~al\mbox{.}(2021)]%
        {austin2021program}
\bibfield{author}{\bibinfo{person}{Jacob Austin}, \bibinfo{person}{Augustus Odena}, \bibinfo{person}{Maxwell Nye}, \bibinfo{person}{Maarten Bosma}, \bibinfo{person}{Henryk Michalewski}, \bibinfo{person}{David Dohan}, \bibinfo{person}{Ellen Jiang}, \bibinfo{person}{Carrie Cai}, \bibinfo{person}{Michael Terry}, \bibinfo{person}{Quoc Le}, {et~al\mbox{.}}} \bibinfo{year}{2021}\natexlab{}.
\newblock \showarticletitle{Program synthesis with large language models}.
\newblock \bibinfo{journal}{\emph{arXiv preprint arXiv:2108.07732}} (\bibinfo{year}{2021}).
\newblock


\bibitem[Chen et~al\mbox{.}({[n.\,d.]})]%
        {chencodet}
\bibfield{author}{\bibinfo{person}{Bei Chen}, \bibinfo{person}{Fengji Zhang}, \bibinfo{person}{Anh Nguyen}, \bibinfo{person}{Daoguang Zan}, \bibinfo{person}{Zeqi Lin}, \bibinfo{person}{Jian-Guang Lou}, {and} \bibinfo{person}{Weizhu Chen}.} \bibinfo{year}{[n.\,d.]}\natexlab{}.
\newblock \showarticletitle{CodeT: Code Generation with Generated Tests}. In \bibinfo{booktitle}{\emph{The Eleventh International Conference on Learning Representations}}.
\newblock


\bibitem[Chen et~al\mbox{.}(2025)]%
        {chen2025self}
\bibfield{author}{\bibinfo{person}{Lili Chen}, \bibinfo{person}{Mihir Prabhudesai}, \bibinfo{person}{Katerina Fragkiadaki}, \bibinfo{person}{Hao Liu}, {and} \bibinfo{person}{Deepak Pathak}.} \bibinfo{year}{2025}\natexlab{}.
\newblock \showarticletitle{Self-questioning language models}.
\newblock \bibinfo{journal}{\emph{arXiv preprint arXiv:2508.03682}} (\bibinfo{year}{2025}).
\newblock


\bibitem[Chen et~al\mbox{.}(2024)]%
        {chen2024b4}
\bibfield{author}{\bibinfo{person}{Mouxiang Chen}, \bibinfo{person}{Zhongxin Liu}, \bibinfo{person}{He Tao}, \bibinfo{person}{Yusu Hong}, \bibinfo{person}{David Lo}, \bibinfo{person}{Xin Xia}, {and} \bibinfo{person}{Jianling Sun}.} \bibinfo{year}{2024}\natexlab{}.
\newblock \showarticletitle{B4: Towards optimal assessment of plausible code solutions with plausible tests}. In \bibinfo{booktitle}{\emph{Proceedings of the 39th IEEE/ACM International Conference on Automated Software Engineering}}. \bibinfo{pages}{1693--1705}.
\newblock


\bibitem[Du et~al\mbox{.}(2023)]%
        {du2023classeval}
\bibfield{author}{\bibinfo{person}{Xueying Du}, \bibinfo{person}{Mingwei Liu}, \bibinfo{person}{Kaixin Wang}, \bibinfo{person}{Hanlin Wang}, \bibinfo{person}{Junwei Liu}, \bibinfo{person}{Yixuan Chen}, \bibinfo{person}{Jiayi Feng}, \bibinfo{person}{Chaofeng Sha}, \bibinfo{person}{Xin Peng}, {and} \bibinfo{person}{Yiling Lou}.} \bibinfo{year}{2023}\natexlab{}.
\newblock \showarticletitle{Classeval: A manually-crafted benchmark for evaluating llms on class-level code generation}.
\newblock \bibinfo{journal}{\emph{arXiv preprint arXiv:2308.01861}} (\bibinfo{year}{2023}).
\newblock


\bibitem[Gorinski et~al\mbox{.}(2023)]%
        {gorinski2023automatic}
\bibfield{author}{\bibinfo{person}{Philip Gorinski}, \bibinfo{person}{Matthieu Zimmer}, \bibinfo{person}{Gerasimos Lampouras}, \bibinfo{person}{Derrick Goh~Xin Deik}, {and} \bibinfo{person}{Ignacio Iacobacci}.} \bibinfo{year}{2023}\natexlab{}.
\newblock \showarticletitle{Automatic unit test data generation and actor-critic reinforcement learning for code synthesis}. In \bibinfo{booktitle}{\emph{Findings of the Association for Computational Linguistics: EMNLP 2023}}. \bibinfo{pages}{370--384}.
\newblock


\bibitem[Guo et~al\mbox{.}(2025)]%
        {guo2025deepseek}
\bibfield{author}{\bibinfo{person}{Daya Guo}, \bibinfo{person}{Dejian Yang}, \bibinfo{person}{Haowei Zhang}, \bibinfo{person}{Junxiao Song}, \bibinfo{person}{Ruoyu Zhang}, \bibinfo{person}{Runxin Xu}, \bibinfo{person}{Qihao Zhu}, \bibinfo{person}{Shirong Ma}, \bibinfo{person}{Peiyi Wang}, \bibinfo{person}{Xiao Bi}, {et~al\mbox{.}}} \bibinfo{year}{2025}\natexlab{}.
\newblock \showarticletitle{Deepseek-r1: Incentivizing reasoning capability in llms via reinforcement learning}.
\newblock \bibinfo{journal}{\emph{arXiv preprint arXiv:2501.12948}} (\bibinfo{year}{2025}).
\newblock


\bibitem[Hendrycks et~al\mbox{.}(2021b)]%
        {hendrycks2measuring}
\bibfield{author}{\bibinfo{person}{Dan Hendrycks}, \bibinfo{person}{Steven Basart}, \bibinfo{person}{Saurav Kadavath}, \bibinfo{person}{Mantas Mazeika}, \bibinfo{person}{Akul Arora}, \bibinfo{person}{Ethan Guo}, \bibinfo{person}{Collin Burns}, \bibinfo{person}{Samir Puranik}, \bibinfo{person}{Horace He}, \bibinfo{person}{Dawn Song}, {et~al\mbox{.}}} \bibinfo{year}{2021}\natexlab{b}.
\newblock \showarticletitle{Measuring Coding Challenge Competence With APPS}. In \bibinfo{booktitle}{\emph{Thirty-fifth Conference on Neural Information Processing Systems Datasets and Benchmarks Track (Round 2)}}.
\newblock


\bibitem[Hendrycks et~al\mbox{.}(2021a)]%
        {hendrycksapps2021}
\bibfield{author}{\bibinfo{person}{Dan Hendrycks}, \bibinfo{person}{Steven Basart}, \bibinfo{person}{Saurav Kadavath}, \bibinfo{person}{Mantas Mazeika}, \bibinfo{person}{Akul Arora}, \bibinfo{person}{Ethan Guo}, \bibinfo{person}{Collin Burns}, \bibinfo{person}{Samir Puranik}, \bibinfo{person}{Horace He}, \bibinfo{person}{Dawn Song}, {and} \bibinfo{person}{Jacob Steinhardt}.} \bibinfo{year}{2021}\natexlab{a}.
\newblock \showarticletitle{Measuring Coding Challenge Competence With APPS}.
\newblock \bibinfo{journal}{\emph{NeurIPS}} (\bibinfo{year}{2021}).
\newblock


\bibitem[Huang et~al\mbox{.}(2025)]%
        {huang2025r}
\bibfield{author}{\bibinfo{person}{Chengsong Huang}, \bibinfo{person}{Wenhao Yu}, \bibinfo{person}{Xiaoyang Wang}, \bibinfo{person}{Hongming Zhang}, \bibinfo{person}{Zongxia Li}, \bibinfo{person}{Ruosen Li}, \bibinfo{person}{Jiaxin Huang}, \bibinfo{person}{Haitao Mi}, {and} \bibinfo{person}{Dong Yu}.} \bibinfo{year}{2025}\natexlab{}.
\newblock \showarticletitle{R-zero: Self-evolving reasoning llm from zero data}.
\newblock \bibinfo{journal}{\emph{arXiv preprint arXiv:2508.05004}} (\bibinfo{year}{2025}).
\newblock


\bibitem[Hui et~al\mbox{.}(2024)]%
        {hui2024qwen2}
\bibfield{author}{\bibinfo{person}{Binyuan Hui}, \bibinfo{person}{Jian Yang}, \bibinfo{person}{Zeyu Cui}, \bibinfo{person}{Jiaxi Yang}, \bibinfo{person}{Dayiheng Liu}, \bibinfo{person}{Lei Zhang}, \bibinfo{person}{Tianyu Liu}, \bibinfo{person}{Jiajun Zhang}, \bibinfo{person}{Bowen Yu}, \bibinfo{person}{Keming Lu}, {et~al\mbox{.}}} \bibinfo{year}{2024}\natexlab{}.
\newblock \showarticletitle{Qwen2. 5-coder technical report}.
\newblock \bibinfo{journal}{\emph{arXiv preprint arXiv:2409.12186}} (\bibinfo{year}{2024}).
\newblock


\bibitem[Inala et~al\mbox{.}(2022)]%
        {inala2022fault}
\bibfield{author}{\bibinfo{person}{Jeevana~Priya Inala}, \bibinfo{person}{Chenglong Wang}, \bibinfo{person}{Mei Yang}, \bibinfo{person}{Andres Codas}, \bibinfo{person}{Mark Encarnaci{\'o}n}, \bibinfo{person}{Shuvendu Lahiri}, \bibinfo{person}{Madanlal Musuvathi}, {and} \bibinfo{person}{Jianfeng Gao}.} \bibinfo{year}{2022}\natexlab{}.
\newblock \showarticletitle{Fault-aware neural code rankers}.
\newblock \bibinfo{journal}{\emph{Advances in Neural Information Processing Systems}}  \bibinfo{volume}{35} (\bibinfo{year}{2022}), \bibinfo{pages}{13419--13432}.
\newblock


\bibitem[Jain et~al\mbox{.}(2025)]%
        {jain2025livecodebench}
\bibfield{author}{\bibinfo{person}{Naman Jain}, \bibinfo{person}{King Han}, \bibinfo{person}{Alex Gu}, \bibinfo{person}{Wen{-}Ding Li}, \bibinfo{person}{Fanjia Yan}, \bibinfo{person}{Tianjun Zhang}, \bibinfo{person}{Sida Wang}, \bibinfo{person}{Armando Solar{-}Lezama}, \bibinfo{person}{Koushik Sen}, {and} \bibinfo{person}{Ion Stoica}.} \bibinfo{year}{2025}\natexlab{}.
\newblock \showarticletitle{LiveCodeBench: Holistic and Contamination Free Evaluation of Large Language Models for Code}. In \bibinfo{booktitle}{\emph{The Thirteenth International Conference on Learning Representations, {ICLR} 2025, Singapore, April 24-28, 2025}}. \bibinfo{publisher}{OpenReview.net}.
\newblock
\urldef\tempurl%
\url{https://openreview.net/forum?id=chfJJYC3iL}
\showURL{%
\tempurl}


\bibitem[Jang et~al\mbox{.}({[n.\,d.]})]%
        {jang2025self}
\bibfield{author}{\bibinfo{person}{Hyosoon Jang}, \bibinfo{person}{Yunhui Jang}, \bibinfo{person}{Sungjae Lee}, \bibinfo{person}{Jungseul Ok}, {and} \bibinfo{person}{Sungsoo Ahn}.} \bibinfo{year}{[n.\,d.]}\natexlab{}.
\newblock \showarticletitle{Self-training large language models with confident reasoning}.
\newblock  (\bibinfo{year}{[n.\,d.]}).
\newblock


\bibitem[Koren et~al\mbox{.}(2009)]%
        {koren2009matrix}
\bibfield{author}{\bibinfo{person}{Yehuda Koren}, \bibinfo{person}{Robert Bell}, {and} \bibinfo{person}{Chris Volinsky}.} \bibinfo{year}{2009}\natexlab{}.
\newblock \showarticletitle{Matrix factorization techniques for recommender systems}.
\newblock \bibinfo{journal}{\emph{Computer}} \bibinfo{volume}{42}, \bibinfo{number}{8} (\bibinfo{year}{2009}), \bibinfo{pages}{30--37}.
\newblock


\bibitem[Laban et~al\mbox{.}(2025)]%
        {laban2025llms}
\bibfield{author}{\bibinfo{person}{Philippe Laban}, \bibinfo{person}{Hiroaki Hayashi}, \bibinfo{person}{Yingbo Zhou}, {and} \bibinfo{person}{Jennifer Neville}.} \bibinfo{year}{2025}\natexlab{}.
\newblock \showarticletitle{Llms get lost in multi-turn conversation}.
\newblock \bibinfo{journal}{\emph{arXiv preprint arXiv:2505.06120}} (\bibinfo{year}{2025}).
\newblock


\bibitem[Lam(2020)]%
        {lam2020high}
\bibfield{author}{\bibinfo{person}{Clifford Lam}.} \bibinfo{year}{2020}\natexlab{}.
\newblock \showarticletitle{High-dimensional covariance matrix estimation}.
\newblock \bibinfo{journal}{\emph{Wiley Interdisciplinary reviews: computational statistics}} \bibinfo{volume}{12}, \bibinfo{number}{2} (\bibinfo{year}{2020}), \bibinfo{pages}{e1485}.
\newblock


\bibitem[Le et~al\mbox{.}(2024)]%
        {lecodechain}
\bibfield{author}{\bibinfo{person}{Hung Le}, \bibinfo{person}{Hailin Chen}, \bibinfo{person}{Amrita Saha}, \bibinfo{person}{Akash Gokul}, \bibinfo{person}{Doyen Sahoo}, {and} \bibinfo{person}{Shafiq Joty}.} \bibinfo{year}{2024}\natexlab{}.
\newblock \showarticletitle{CodeChain: Towards Modular Code Generation Through Chain of Self-revisions with Representative Sub-modules}. In \bibinfo{booktitle}{\emph{The Twelfth International Conference on Learning Representations}}.
\newblock


\bibitem[Li et~al\mbox{.}(2022)]%
        {li2022competition}
\bibfield{author}{\bibinfo{person}{Yujia Li}, \bibinfo{person}{David Choi}, \bibinfo{person}{Junyoung Chung}, \bibinfo{person}{Nate Kushman}, \bibinfo{person}{Julian Schrittwieser}, \bibinfo{person}{R{\'e}mi Leblond}, \bibinfo{person}{Tom Eccles}, \bibinfo{person}{James Keeling}, \bibinfo{person}{Felix Gimeno}, \bibinfo{person}{Agustin Dal~Lago}, {et~al\mbox{.}}} \bibinfo{year}{2022}\natexlab{}.
\newblock \showarticletitle{Competition-level code generation with alphacode}.
\newblock \bibinfo{journal}{\emph{Science}} \bibinfo{volume}{378}, \bibinfo{number}{6624} (\bibinfo{year}{2022}), \bibinfo{pages}{1092--1097}.
\newblock


\bibitem[Lin et~al\mbox{.}(2025)]%
        {lin2025learning}
\bibfield{author}{\bibinfo{person}{Zi Lin}, \bibinfo{person}{Sheng Shen}, \bibinfo{person}{Jingbo Shang}, \bibinfo{person}{Jason Weston}, {and} \bibinfo{person}{Yixin Nie}.} \bibinfo{year}{2025}\natexlab{}.
\newblock \showarticletitle{Learning to solve and verify: A self-play framework for code and test generation}.
\newblock \bibinfo{journal}{\emph{arXiv preprint arXiv:2502.14948}} (\bibinfo{year}{2025}).
\newblock


\bibitem[Liu et~al\mbox{.}(2024a)]%
        {liu2024your}
\bibfield{author}{\bibinfo{person}{Jiawei Liu}, \bibinfo{person}{Chunqiu~Steven Xia}, \bibinfo{person}{Yuyao Wang}, {and} \bibinfo{person}{Lingming Zhang}.} \bibinfo{year}{2024}\natexlab{a}.
\newblock \showarticletitle{Is your code generated by chatgpt really correct? rigorous evaluation of large language models for code generation}.
\newblock \bibinfo{journal}{\emph{Advances in Neural Information Processing Systems}}  \bibinfo{volume}{36} (\bibinfo{year}{2024}).
\newblock


\bibitem[Liu et~al\mbox{.}(2024b)]%
        {liu2024dstc}
\bibfield{author}{\bibinfo{person}{Zhihan Liu}, \bibinfo{person}{Shenao Zhang}, \bibinfo{person}{Yongfei Liu}, \bibinfo{person}{Boyi Liu}, \bibinfo{person}{Yingxiang Yang}, {and} \bibinfo{person}{Zhaoran Wang}.} \bibinfo{year}{2024}\natexlab{b}.
\newblock \showarticletitle{Dstc: Direct preference learning with only self-generated tests and code to improve code lms}.
\newblock \bibinfo{journal}{\emph{arXiv preprint arXiv:2411.13611}} (\bibinfo{year}{2024}).
\newblock


\bibitem[Loshchilov and Hutter(2017)]%
        {loshchilov2017sgdr}
\bibfield{author}{\bibinfo{person}{Ilya Loshchilov} {and} \bibinfo{person}{Frank Hutter}.} \bibinfo{year}{2017}\natexlab{}.
\newblock \showarticletitle{SGDR: Stochastic Gradient Descent with Warm Restarts}. In \bibinfo{booktitle}{\emph{International Conference on Learning Representations}}.
\newblock


\bibitem[Ma et~al\mbox{.}(2025)]%
        {ma2025unitcoder}
\bibfield{author}{\bibinfo{person}{Yichuan Ma}, \bibinfo{person}{Yunfan Shao}, \bibinfo{person}{Peiji Li}, \bibinfo{person}{Demin Song}, \bibinfo{person}{Qipeng Guo}, \bibinfo{person}{Linyang Li}, \bibinfo{person}{Xipeng Qiu}, {and} \bibinfo{person}{Kai Chen}.} \bibinfo{year}{2025}\natexlab{}.
\newblock \showarticletitle{UnitCoder: Scalable Code Synthesis from Pre-training Corpora}. In \bibinfo{booktitle}{\emph{Proceedings of the 2025 Conference on Empirical Methods in Natural Language Processing}}. \bibinfo{pages}{5623--5641}.
\newblock


\bibitem[Nichol and Dhariwal(2021)]%
        {nichol2021improved}
\bibfield{author}{\bibinfo{person}{Alexander~Quinn Nichol} {and} \bibinfo{person}{Prafulla Dhariwal}.} \bibinfo{year}{2021}\natexlab{}.
\newblock \showarticletitle{Improved denoising diffusion probabilistic models}. In \bibinfo{booktitle}{\emph{International conference on machine learning}}. PMLR, \bibinfo{pages}{8162--8171}.
\newblock


\bibitem[Olausson et~al\mbox{.}(2023)]%
        {olausson2023self}
\bibfield{author}{\bibinfo{person}{Theo~X Olausson}, \bibinfo{person}{Jeevana~Priya Inala}, \bibinfo{person}{Chenglong Wang}, \bibinfo{person}{Jianfeng Gao}, {and} \bibinfo{person}{Armando Solar-Lezama}.} \bibinfo{year}{2023}\natexlab{}.
\newblock \showarticletitle{Is Self-Repair a Silver Bullet for Code Generation?}. In \bibinfo{booktitle}{\emph{The Twelfth International Conference on Learning Representations}}.
\newblock


\bibitem[Penedo et~al\mbox{.}(2025)]%
        {penedo2025codeforces}
\bibfield{author}{\bibinfo{person}{Guilherme Penedo}, \bibinfo{person}{Anton Lozhkov}, \bibinfo{person}{Hynek Kydlíček}, \bibinfo{person}{Loubna~Ben Allal}, \bibinfo{person}{Edward Beeching}, \bibinfo{person}{Agustín~Piqueres Lajarín}, \bibinfo{person}{Quentin Gallouédec}, \bibinfo{person}{Nathan Habib}, \bibinfo{person}{Lewis Tunstall}, {and} \bibinfo{person}{Leandro von Werra}.} \bibinfo{year}{2025}\natexlab{}.
\newblock \bibinfo{title}{CodeForces}.
\newblock \bibinfo{howpublished}{\url{https://huggingface.co/datasets/open-r1/codeforces}}.
\newblock


\bibitem[Prabhudesai et~al\mbox{.}(2025)]%
        {prabhudesai2025maximizing}
\bibfield{author}{\bibinfo{person}{Mihir Prabhudesai}, \bibinfo{person}{Lili Chen}, \bibinfo{person}{Alex Ippoliti}, \bibinfo{person}{Katerina Fragkiadaki}, \bibinfo{person}{Hao Liu}, {and} \bibinfo{person}{Deepak Pathak}.} \bibinfo{year}{2025}\natexlab{}.
\newblock \showarticletitle{Maximizing confidence alone improves reasoning}.
\newblock \bibinfo{journal}{\emph{arXiv preprint arXiv:2505.22660}} (\bibinfo{year}{2025}).
\newblock


\bibitem[Prasad et~al\mbox{.}(2025)]%
        {prasad2025learning}
\bibfield{author}{\bibinfo{person}{Archiki Prasad}, \bibinfo{person}{Elias Stengel-Eskin}, \bibinfo{person}{Justin Chih-Yao Chen}, \bibinfo{person}{Zaid Khan}, {and} \bibinfo{person}{Mohit Bansal}.} \bibinfo{year}{2025}\natexlab{}.
\newblock \showarticletitle{Learning to generate unit tests for automated debugging}.
\newblock \bibinfo{journal}{\emph{arXiv preprint arXiv:2502.01619}} (\bibinfo{year}{2025}).
\newblock


\bibitem[Roy and Vetterli(2007)]%
        {roy2007effective}
\bibfield{author}{\bibinfo{person}{Olivier Roy} {and} \bibinfo{person}{Martin Vetterli}.} \bibinfo{year}{2007}\natexlab{}.
\newblock \showarticletitle{The effective rank: A measure of effective dimensionality}. In \bibinfo{booktitle}{\emph{2007 15th European signal processing conference}}. IEEE, \bibinfo{pages}{606--610}.
\newblock


\bibitem[Shafayat et~al\mbox{.}(2025)]%
        {shafayat2025can}
\bibfield{author}{\bibinfo{person}{Sheikh Shafayat}, \bibinfo{person}{Fahim Tajwar}, \bibinfo{person}{Ruslan Salakhutdinov}, \bibinfo{person}{Jeff Schneider}, {and} \bibinfo{person}{Andrea Zanette}.} \bibinfo{year}{2025}\natexlab{}.
\newblock \showarticletitle{Can Large Reasoning Models Self-Train?}
\newblock \bibinfo{journal}{\emph{arXiv preprint arXiv:2505.21444}} (\bibinfo{year}{2025}).
\newblock


\bibitem[Shao et~al\mbox{.}(2024)]%
        {shao2024deepseekmath}
\bibfield{author}{\bibinfo{person}{Zhihong Shao}, \bibinfo{person}{Peiyi Wang}, \bibinfo{person}{Qihao Zhu}, \bibinfo{person}{Runxin Xu}, \bibinfo{person}{Junxiao Song}, \bibinfo{person}{Xiao Bi}, \bibinfo{person}{Haowei Zhang}, \bibinfo{person}{Mingchuan Zhang}, \bibinfo{person}{YK Li}, \bibinfo{person}{Yang Wu}, {et~al\mbox{.}}} \bibinfo{year}{2024}\natexlab{}.
\newblock \showarticletitle{Deepseekmath: Pushing the limits of mathematical reasoning in open language models}.
\newblock \bibinfo{journal}{\emph{arXiv preprint arXiv:2402.03300}} (\bibinfo{year}{2024}).
\newblock


\bibitem[Sheng et~al\mbox{.}(2025)]%
        {sheng2025hybridflow}
\bibfield{author}{\bibinfo{person}{Guangming Sheng}, \bibinfo{person}{Chi Zhang}, \bibinfo{person}{Zilingfeng Ye}, \bibinfo{person}{Xibin Wu}, \bibinfo{person}{Wang Zhang}, \bibinfo{person}{Ru Zhang}, \bibinfo{person}{Yanghua Peng}, \bibinfo{person}{Haibin Lin}, {and} \bibinfo{person}{Chuan Wu}.} \bibinfo{year}{2025}\natexlab{}.
\newblock \showarticletitle{Hybridflow: A flexible and efficient rlhf framework}. In \bibinfo{booktitle}{\emph{Proceedings of the Twentieth European Conference on Computer Systems}}. \bibinfo{pages}{1279--1297}.
\newblock


\bibitem[Shi et~al\mbox{.}(2022)]%
        {shi2022natural}
\bibfield{author}{\bibinfo{person}{Freda Shi}, \bibinfo{person}{Daniel Fried}, \bibinfo{person}{Marjan Ghazvininejad}, \bibinfo{person}{Luke Zettlemoyer}, {and} \bibinfo{person}{Sida~I Wang}.} \bibinfo{year}{2022}\natexlab{}.
\newblock \showarticletitle{Natural language to code translation with execution}. In \bibinfo{booktitle}{\emph{Proceedings of the 2022 Conference on Empirical Methods in Natural Language Processing}}. \bibinfo{pages}{3533--3546}.
\newblock


\bibitem[Wang et~al\mbox{.}(2025)]%
        {wang2025co}
\bibfield{author}{\bibinfo{person}{Yinjie Wang}, \bibinfo{person}{Ling Yang}, \bibinfo{person}{Ye Tian}, \bibinfo{person}{Ke Shen}, {and} \bibinfo{person}{Mengdi Wang}.} \bibinfo{year}{2025}\natexlab{}.
\newblock \showarticletitle{Co-evolving llm coder and unit tester via reinforcement learning}.
\newblock \bibinfo{journal}{\emph{arXiv preprint arXiv:2506.03136}} (\bibinfo{year}{2025}).
\newblock


\bibitem[Wang et~al\mbox{.}(2023)]%
        {wang2023execution}
\bibfield{author}{\bibinfo{person}{Zhiruo Wang}, \bibinfo{person}{Shuyan Zhou}, \bibinfo{person}{Daniel Fried}, {and} \bibinfo{person}{Graham Neubig}.} \bibinfo{year}{2023}\natexlab{}.
\newblock \showarticletitle{Execution-based evaluation for open-domain code generation}. In \bibinfo{booktitle}{\emph{Findings of the Association for Computational Linguistics: EMNLP 2023}}. \bibinfo{pages}{1271--1290}.
\newblock


\bibitem[Wen et~al\mbox{.}(2025)]%
        {wen2025light}
\bibfield{author}{\bibinfo{person}{Liang Wen}, \bibinfo{person}{Yunke Cai}, \bibinfo{person}{Fenrui Xiao}, \bibinfo{person}{Xin He}, \bibinfo{person}{Qi An}, \bibinfo{person}{Zhenyu Duan}, \bibinfo{person}{Yimin Du}, \bibinfo{person}{Junchen Liu}, \bibinfo{person}{Tanglifu Tanglifu}, \bibinfo{person}{Xiaowei Lv}, {et~al\mbox{.}}} \bibinfo{year}{2025}\natexlab{}.
\newblock \showarticletitle{Light-r1: Curriculum sft, dpo and rl for long cot from scratch and beyond}. In \bibinfo{booktitle}{\emph{Proceedings of the 63rd Annual Meeting of the Association for Computational Linguistics (Volume 6: Industry Track)}}. \bibinfo{pages}{318--327}.
\newblock


\bibitem[Xiong et~al\mbox{.}(2024)]%
        {xiong2024program}
\bibfield{author}{\bibinfo{person}{Weimin Xiong}, \bibinfo{person}{Yiwen Guo}, {and} \bibinfo{person}{Hao Chen}.} \bibinfo{year}{2024}\natexlab{}.
\newblock \showarticletitle{The program testing ability of large language models for code}. In \bibinfo{booktitle}{\emph{Proceedings of the 2024 Conference on Empirical Methods in Natural Language Processing: Industry Track}}. \bibinfo{pages}{23--34}.
\newblock


\bibitem[Xu et~al\mbox{.}(2025)]%
        {xu2025kodcode}
\bibfield{author}{\bibinfo{person}{Zhangchen Xu}, \bibinfo{person}{Yang Liu}, \bibinfo{person}{Yueqin Yin}, \bibinfo{person}{Mingyuan Zhou}, {and} \bibinfo{person}{Radha Poovendran}.} \bibinfo{year}{2025}\natexlab{}.
\newblock \showarticletitle{Kodcode: A diverse, challenging, and verifiable synthetic dataset for coding}.
\newblock \bibinfo{journal}{\emph{arXiv preprint arXiv:2503.02951}} (\bibinfo{year}{2025}).
\newblock


\bibitem[Yang et~al\mbox{.}(2025)]%
        {yang2025qwen3}
\bibfield{author}{\bibinfo{person}{An Yang}, \bibinfo{person}{Anfeng Li}, \bibinfo{person}{Baosong Yang}, \bibinfo{person}{Beichen Zhang}, \bibinfo{person}{Binyuan Hui}, \bibinfo{person}{Bo Zheng}, \bibinfo{person}{Bowen Yu}, \bibinfo{person}{Chang Gao}, \bibinfo{person}{Chengen Huang}, \bibinfo{person}{Chenxu Lv}, {et~al\mbox{.}}} \bibinfo{year}{2025}\natexlab{}.
\newblock \showarticletitle{Qwen3 technical report}.
\newblock \bibinfo{journal}{\emph{arXiv preprint arXiv:2505.09388}} (\bibinfo{year}{2025}).
\newblock


\bibitem[Yang et~al\mbox{.}(2024)]%
        {yang2025qwen2}
\bibfield{author}{\bibinfo{person}{An Yang}, \bibinfo{person}{Baosong Yang}, \bibinfo{person}{Beichen Zhang}, \bibinfo{person}{Binyuan Hui}, \bibinfo{person}{Bo Zheng}, \bibinfo{person}{Bowen Yu}, \bibinfo{person}{Chengyuan Li}, \bibinfo{person}{Dayiheng Liu}, \bibinfo{person}{Fei Huang}, \bibinfo{person}{Haoran Wei}, \bibinfo{person}{Huan Lin}, \bibinfo{person}{Jian Yang}, \bibinfo{person}{Jianhong Tu}, \bibinfo{person}{Jianwei Zhang}, \bibinfo{person}{Jianxin Yang}, \bibinfo{person}{Jiaxi Yang}, \bibinfo{person}{Jingren Zhou}, \bibinfo{person}{Junyang Lin}, \bibinfo{person}{Kai Dang}, \bibinfo{person}{Keming Lu}, \bibinfo{person}{Keqin Bao}, \bibinfo{person}{Kexin Yang}, \bibinfo{person}{Le Yu}, \bibinfo{person}{Mei Li}, \bibinfo{person}{Mingfeng Xue}, \bibinfo{person}{Pei Zhang}, \bibinfo{person}{Qin Zhu}, \bibinfo{person}{Rui Men}, \bibinfo{person}{Runji Lin}, \bibinfo{person}{Tianhao Li}, \bibinfo{person}{Tingyu Xia}, \bibinfo{person}{Xingzhang Ren}, \bibinfo{person}{Xuancheng Ren}, \bibinfo{person}{Yang
  Fan}, \bibinfo{person}{Yang Su}, \bibinfo{person}{Yichang Zhang}, \bibinfo{person}{Yu Wan}, \bibinfo{person}{Yuqiong Liu}, \bibinfo{person}{Zeyu Cui}, \bibinfo{person}{Zhenru Zhang}, {and} \bibinfo{person}{Zihan Qiu}.} \bibinfo{year}{2024}\natexlab{}.
\newblock \showarticletitle{Qwen2.5 Technical Report}.
\newblock \bibinfo{journal}{\emph{CoRR}}  \bibinfo{volume}{abs/2412.15115} (\bibinfo{year}{2024}).
\newblock
\showeprint[arXiv]{2412.15115}
\href{https://doi.org/10.48550/ARXIV.2412.15115}{doi:\nolinkurl{10.48550/ARXIV.2412.15115}}


\bibitem[Yu et~al\mbox{.}(2024)]%
        {yu2024codereval}
\bibfield{author}{\bibinfo{person}{Hao Yu}, \bibinfo{person}{Bo Shen}, \bibinfo{person}{Dezhi Ran}, \bibinfo{person}{Jiaxin Zhang}, \bibinfo{person}{Qi Zhang}, \bibinfo{person}{Yuchi Ma}, \bibinfo{person}{Guangtai Liang}, \bibinfo{person}{Ying Li}, \bibinfo{person}{Qianxiang Wang}, {and} \bibinfo{person}{Tao Xie}.} \bibinfo{year}{2024}\natexlab{}.
\newblock \showarticletitle{Codereval: A benchmark of pragmatic code generation with generative pre-trained models}. In \bibinfo{booktitle}{\emph{Proceedings of the 46th IEEE/ACM International Conference on Software Engineering}}. \bibinfo{pages}{1--12}.
\newblock


\bibitem[Yu et~al\mbox{.}(2025)]%
        {yu2025dapo}
\bibfield{author}{\bibinfo{person}{Qiying Yu}, \bibinfo{person}{Zheng Zhang}, \bibinfo{person}{Ruofei Zhu}, \bibinfo{person}{Yufeng Yuan}, \bibinfo{person}{Xiaochen Zuo}, \bibinfo{person}{Yu Yue}, \bibinfo{person}{Weinan Dai}, \bibinfo{person}{Tiantian Fan}, \bibinfo{person}{Gaohong Liu}, \bibinfo{person}{Lingjun Liu}, {et~al\mbox{.}}} \bibinfo{year}{2025}\natexlab{}.
\newblock \showarticletitle{Dapo: An open-source llm reinforcement learning system at scale}.
\newblock \bibinfo{journal}{\emph{arXiv preprint arXiv:2503.14476}} (\bibinfo{year}{2025}).
\newblock


\bibitem[Zawalski et~al\mbox{.}(2021)]%
        {zawalski2021off}
\bibfield{author}{\bibinfo{person}{Micha{\l} Zawalski}, \bibinfo{person}{B{\l}a{\.z}ej Osi{\'n}ski}, \bibinfo{person}{Henryk Michalewski}, {and} \bibinfo{person}{Piotr Mi{\l}o{\'s}}.} \bibinfo{year}{2021}\natexlab{}.
\newblock \showarticletitle{Off-policy correction for multi-agent reinforcement learning}.
\newblock \bibinfo{journal}{\emph{arXiv preprint arXiv:2111.11229}} (\bibinfo{year}{2021}).
\newblock


\bibitem[Zeng et~al\mbox{.}(2025)]%
        {zeng2025acecoder}
\bibfield{author}{\bibinfo{person}{Huaye Zeng}, \bibinfo{person}{Dongfu Jiang}, \bibinfo{person}{Haozhe Wang}, \bibinfo{person}{Ping Nie}, \bibinfo{person}{Xiaotong Chen}, {and} \bibinfo{person}{Wenhu Chen}.} \bibinfo{year}{2025}\natexlab{}.
\newblock \showarticletitle{Acecoder: Acing coder rl via automated test-case synthesis}.
\newblock \bibinfo{journal}{\emph{arXiv preprint arXiv:2502.01718}} (\bibinfo{year}{2025}).
\newblock


\bibitem[Zhang et~al\mbox{.}(2025a)]%
        {zhang2025codedpo}
\bibfield{author}{\bibinfo{person}{Kechi Zhang}, \bibinfo{person}{Ge Li}, \bibinfo{person}{Yihong Dong}, \bibinfo{person}{Jingjing Xu}, \bibinfo{person}{Jun Zhang}, \bibinfo{person}{Jing Su}, \bibinfo{person}{Yongfei Liu}, {and} \bibinfo{person}{Zhi Jin}.} \bibinfo{year}{2025}\natexlab{a}.
\newblock \showarticletitle{Codedpo: Aligning code models with self generated and verified source code}. In \bibinfo{booktitle}{\emph{Proceedings of the 63rd Annual Meeting of the Association for Computational Linguistics (Volume 1: Long Papers)}}. \bibinfo{pages}{15854--15871}.
\newblock


\bibitem[Zhang et~al\mbox{.}(2025b)]%
        {zhang2025right}
\bibfield{author}{\bibinfo{person}{Qingyang Zhang}, \bibinfo{person}{Haitao Wu}, \bibinfo{person}{Changqing Zhang}, \bibinfo{person}{Peilin Zhao}, {and} \bibinfo{person}{Yatao Bian}.} \bibinfo{year}{2025}\natexlab{b}.
\newblock \showarticletitle{Right question is already half the answer: Fully unsupervised llm reasoning incentivization}.
\newblock \bibinfo{journal}{\emph{arXiv preprint arXiv:2504.05812}} (\bibinfo{year}{2025}).
\newblock


\bibitem[Zhang et~al\mbox{.}(2023)]%
        {zhang2023coder}
\bibfield{author}{\bibinfo{person}{Tianyi Zhang}, \bibinfo{person}{Tao Yu}, \bibinfo{person}{Tatsunori Hashimoto}, \bibinfo{person}{Mike Lewis}, \bibinfo{person}{Wen-tau Yih}, \bibinfo{person}{Daniel Fried}, {and} \bibinfo{person}{Sida Wang}.} \bibinfo{year}{2023}\natexlab{}.
\newblock \showarticletitle{Coder reviewer reranking for code generation}. In \bibinfo{booktitle}{\emph{International Conference on Machine Learning}}. PMLR, \bibinfo{pages}{41832--41846}.
\newblock


\bibitem[Zhang et~al\mbox{.}(2025c)]%
        {zhang2025unseen}
\bibfield{author}{\bibinfo{person}{Yuanliang Zhang}, \bibinfo{person}{Yifan Xie}, \bibinfo{person}{Shanshan Lit}, \bibinfo{person}{Ke Liu}, \bibinfo{person}{Chong Wang}, \bibinfo{person}{Zhouyang Jia}, \bibinfo{person}{Xiangbing Huang}, \bibinfo{person}{Jie Song}, \bibinfo{person}{Chaopeng Luo}, \bibinfo{person}{Zhizheng Zheng}, {et~al\mbox{.}}} \bibinfo{year}{2025}\natexlab{c}.
\newblock \showarticletitle{Unseen horizons: Unveiling the real capability of llm code generation beyond the familiar}. In \bibinfo{booktitle}{\emph{2025 IEEE/ACM 47th International Conference on Software Engineering (ICSE)}}. IEEE, \bibinfo{pages}{604--615}.
\newblock


\bibitem[Zhao et~al\mbox{.}(2025)]%
        {zhao2025absolute}
\bibfield{author}{\bibinfo{person}{Andrew Zhao}, \bibinfo{person}{Yiran Wu}, \bibinfo{person}{Yang Yue}, \bibinfo{person}{Tong Wu}, \bibinfo{person}{Quentin Xu}, \bibinfo{person}{Matthieu Lin}, \bibinfo{person}{Shenzhi Wang}, \bibinfo{person}{Qingyun Wu}, \bibinfo{person}{Zilong Zheng}, {and} \bibinfo{person}{Gao Huang}.} \bibinfo{year}{2025}\natexlab{}.
\newblock \showarticletitle{Absolute zero: Reinforced self-play reasoning with zero data}.
\newblock \bibinfo{journal}{\emph{arXiv preprint arXiv:2505.03335}} (\bibinfo{year}{2025}).
\newblock


\bibitem[Zheng et~al\mbox{.}(2025)]%
        {zheng2025group}
\bibfield{author}{\bibinfo{person}{Chujie Zheng}, \bibinfo{person}{Shixuan Liu}, \bibinfo{person}{Mingze Li}, \bibinfo{person}{Xiong-Hui Chen}, \bibinfo{person}{Bowen Yu}, \bibinfo{person}{Chang Gao}, \bibinfo{person}{Kai Dang}, \bibinfo{person}{Yuqiong Liu}, \bibinfo{person}{Rui Men}, \bibinfo{person}{An Yang}, {et~al\mbox{.}}} \bibinfo{year}{2025}\natexlab{}.
\newblock \showarticletitle{Group sequence policy optimization}.
\newblock \bibinfo{journal}{\emph{arXiv preprint arXiv:2507.18071}} (\bibinfo{year}{2025}).
\newblock


\bibitem[Zuo et~al\mbox{.}(2025)]%
        {zuo2025ttrl}
\bibfield{author}{\bibinfo{person}{Yuxin Zuo}, \bibinfo{person}{Kaiyan Zhang}, \bibinfo{person}{Li Sheng}, \bibinfo{person}{Shang Qu}, \bibinfo{person}{Ganqu Cui}, \bibinfo{person}{Xuekai Zhu}, \bibinfo{person}{Haozhan Li}, \bibinfo{person}{Yuchen Zhang}, \bibinfo{person}{Xinwei Long}, \bibinfo{person}{Ermo Hua}, {et~al\mbox{.}}} \bibinfo{year}{2025}\natexlab{}.
\newblock \showarticletitle{Ttrl: Test-time reinforcement learning}.
\newblock \bibinfo{journal}{\emph{arXiv preprint arXiv:2504.16084}} (\bibinfo{year}{2025}).
\newblock


\end{thebibliography}

\appendix

\end{document}